\documentclass[12pt,draftclsnofoot, perreview, onecolumn]{IEEEtran}

\normalsize

\usepackage{amsfonts}
\usepackage{amsmath}
\usepackage{mathrsfs}
\usepackage{amssymb}
\usepackage{booktabs}
\usepackage{balance}
\usepackage{color}
\usepackage{citesort}
\usepackage{enumerate}
\usepackage{graphicx}
\usepackage{mdwmath}
\usepackage{multirow,tabularx}
\usepackage{subfigure}

\ifCLASSINFOpdf
\else
\fi

\ifCLASSOPTIONonecolumn
\newcommand{\figwidth}{0.56\textwidth}
\newcommand{\smallfigwidth}{0.28\textwidth}
\newcommand{\Twofigwidth}{0.85\textwidth}
\fi
\ifCLASSOPTIONtwocolumn
\newcommand{\figwidth}{0.45\textwidth}
\newcommand{\smallfigwidth}{0.21\textwidth}
\newcommand{\Twofigwidth}{0.85\textwidth}
\fi

\makeatletter
    
    \newcommand{\Rmnum}[1]{\expandafter\@slowromancap\romannumeral #1@}
\makeatother

\ifCLASSOPTIONonecolumn
\fi

\begin{document}
%
\title{Performance Comparison of LDPC Block and Spatially Coupled Codes over GF($q$)}

\author{Kechao~Huang,~\IEEEmembership{Student~Member,~IEEE,}
        David~G.~M.~Mitchell,~\IEEEmembership{Member,~IEEE,}
        Lai~Wei,~\IEEEmembership{Student~Member,~IEEE,}
        Xiao~Ma,~\IEEEmembership{Member,~IEEE,}
        and~Daniel~J.~Costello,~Jr.,~\IEEEmembership{Life~Fellow,~IEEE}

        \thanks{This work was partially supported by the Joint Ph.D. Fellowship Program of the China Scholarship Council, the $973$ Program (No. $2012$CB$316100$), the China NSF (No. 61172082), and the U.S. NSF (No. CCF-1161754). This work was performed while K. Huang was visiting the University of Notre Dame. The material in this paper was presented in part at the Information Theory and Applications Workshop, San Diego, CA, Feb. 2014, and in part at the IEEE International Symposium on Information Theory, Honolulu, HI, July 2014.}

        \thanks{K.~Huang and X.~Ma are with the Department of Electronics and Communication Engineering, Sun Yat-sen University, Guangzhou, GD 510006, China~(e-mail:~hkech@mail2.sysu.edu.cn; maxiao@mail.sysu.edu.cn). K.~Huang is also with the Department of Electrical Engineering, University of Notre Dame, Notre Dame, IN 46556, USA.}
        \thanks{D.~G.~M.~Mitchell, L.~Wei, and D.~J.~Costello,~Jr. are with the Department of Electrical Engineering, University of Notre Dame, Notre Dame, IN 46556, USA~(e-mail:~david.mitchell@nd.edu; lwei1@nd.edu; costello.2@nd.edu).}
}
\markboth{IEEE Trans.~Commun. (Submitted Paper)}{}%
\maketitle

\begin{abstract}
In this paper, we compare the finite-length performance of protograph-based spatially coupled low-density parity-check~(SC-LDPC) codes and LDPC block codes~(LDPC-BCs) over GF($q$). In order to reduce computational complexity and latency, a sliding window decoder with a stopping rule based on a soft bit-error-rate~(BER) estimate is used for the $q$-ary SC-LDPC codes. Two regimes are considered: one when the constraint length of $q$-ary SC-LDPC codes is equal to the block length of $q$-ary LDPC-BCs and the other when the two decoding latencies are equal. Simulation results confirm that, in both regimes, $(3,6)$-, $(3,9)$-, and $(3,12)$-regular non-binary SC-LDPC codes can significantly outperform both binary and non-binary LDPC-BCs and binary SC-LDPC codes. Finally, we present a computational complexity comparison of $q$-ary SC-LDPC codes and $q$-ary LDPC-BCs under equal decoding latency and equal decoding performance assumptions.
\ifCLASSOPTIONonecolumn
\vspace{1cm}
\fi
\end{abstract}

\begin{IEEEkeywords}
Decoding latency, LDPC block codes, LDPC convolutional codes, protograph-based codes, $q$-ary LDPC codes, spatially coupled codes.
\end{IEEEkeywords}

\section{Introduction}
Low-density parity-check block codes~(LDPC-BCs)~\cite{Gallager63}, combined with low complexity belief propagation~(BP) decoding algorithms, are a class of capacity-approaching codes with decoding complexity that increases only linearly with block length~\cite{Richardson01}. In~\cite{Gallager63}, in addition to binary LDPC-BCs, Gallager also introduced a class of non-binary LDPC-BCs defined over an arbitrary alphabet size. In~\cite{Davey98}, Davey and MacKay considered LDPC-BCs defined over a finite field GF($q$)$,~q\geqslant 2$, and generalized Gallager's BP decoding algorithm for binary LDPC-BCs to a $q$-ary sum-product algorithm~(QSPA) and demonstrated that $q$-ary LDPC-BCs achieve excellent performance. To reduce decoding complexity, a more efficient QSPA based on the fast Fourier transform~(called FFT-QSPA) was proposed in~\cite{Barnault03}. In addition, extended min-sum~(EMS) algorithms~\cite{Declercq07,Voicila10,Ma12} can be used to further reduce decoding complexity. Due to their excellent decoding performance for short-to-moderate block lengths~\cite{Davey98}, $q$-ary LDPC-BCs have received significant attention in the recent literature~\cite{poulliat08,Zeng08,Zhao13,Dolecek13}.

The convolutional counterpart of LDPC-BCs, called spatially coupled LDPC~(SC-LDPC) codes, was proposed in~\cite{jimenez99}. Analogous to LDPC-BCs, SC-LDPC codes are defined by sparse parity-check matrices, which allow them to be decoded using iterative message-passing algorithms, such as BP decoding algorithms. It was shown in~\cite{lentmaier10} that the BP decoding thresholds of SC-LDPC code ensembles are numerically indistinguishable from the maximum {\em a posteriori}~(MAP) decoding thresholds of underlying regular and irregular LDPC-BC ensembles. Subsequently, it was proven that random SC-LDPC code ensembles exhibit {\em threshold saturation}, i.e., they achieve the MAP thresholds of the underlying LDPC-BCs, on memoryless binary-input symmetric-output channels under BP decoding, which in turn implies that SC-LDPC codes can achieve capacity by increasing the density of the parity-check matrix~\cite{kudekar11,kudekar13}. In~\cite{jimenez99}, a parallel, high-speed, pipeline-decoding architecture for  binary SC-LDPC codes was introduced, and several implementation aspects of the pipeline decoder were discussed in~\cite{Ali08}. However, since capacity approaching performance can require a large number of iterations, the latency and memory requirements of the pipeline decoder, which depend on the number of iterations, may be unacceptably high. In~\cite{iyengar2012windowed}, a sliding window decoding architecture with reduced latency and memory requirements was proposed. This is a variant of the sliding window decoder introduced in~\cite{lentmaier10} for the purpose of iterative decoding threshold analysis. A construction method for $q$-ary SC-LDPC codes was introduced in~\cite{Uchikawa11}, and in~\cite{Piemontese13} the authors proved that the threshold saturation effect proved in~\cite{kudekar11} for binary SC-LDPC codes also holds for $q$-ary SC-LDPC codes on the binary erasure channel~(BEC). Recently, based on numerical techniques, the threshold performance of $q$-ary SC-LDPC codes constructed from protographs~\cite{Thorpe03} with sliding window decoding was presented in~\cite{Wei13,Wei13IT}.

In contrast to~\cite{Wei13,Wei13IT}, in which the authors consider an asymptotic performance analysis of $q$-ary SC-LDPC codes, in this paper we focus on finite-length performance comparisons of protograph-based $q$-ary SC-LDPC codes and $q$-ary LDPC-BCs, assuming transmission over a binary-input additive white Gaussian noise~(BI-AWGN) channel. Due to the large decoding latency of the pipeline-decoding architecture, a sliding window decoder for $q$-ary SC-LDPC codes is considered. In order to reduce computational complexity, a stopping rule based on a soft bit-error-rate~(BER) estimate is applied to the iterative decoding process. Two regimes are considered: one when the constraint length of $q$-ary SC-LDPC codes is equal to the block length of $q$-ary LDPC-BCs and the other when the two decoding latencies are equal. We also investigate the relationship between the protograph lifting factor, the decoding window size, and the decoding performance of $q$-ary SC-LDPC codes when the decoding latency is fixed. Finally, we compare the computational complexity of $q$-ary SC-LDPC codes to $q$-ary LDPC-BCs when either the decoding latency or the decoding performance is fixed.

The paper is structured as follows. In Section~\ref{sec:Protograph}, we give a brief review of protograph-based LDPC-BCs and then describe the construction of protograph-based $q$-ary SC-LDPC codes. In Section~\ref{sec:Decoding}, we describe the pipeline and sliding window decoding architectures and introduce a stopping rule based on a soft BER estimate for $q$-ary SC-LDPC codes. In Section~\ref{sec:CG}, we present a performance comparison of $q$-ary SC-LDPC codes and $q$-ary LDPC-BCs when the constraint length of $q$-ary SC-LDPC codes is equal to the block length of $q$-ary LDPC-BCs, and in Section~\ref{sec:ELC} we compare their performance on the basis of equal decoding latency. Then, in Section~\ref{sec:Complexity}, we compare the computational complexity of $q$-ary SC-LDPC codes and $q$-ary LDPC-BCs under equal decoding latency and equal decoding performance assumptions. Finally, some concluding remarks are given in Section~\ref{sec:Conclusion}.

\section{Protograph-Based LDPC Codes over GF($q$)}\label{sec:Protograph}
\subsection{LDPC-BCs over GF($q$)}\label{subsec:LDPC-BC}

\begin{figure}
    \center
    \includegraphics[angle=270, clip, width=\smallfigwidth]{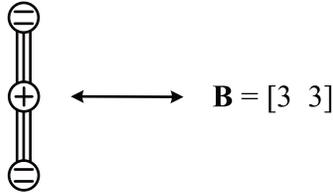}
    \caption{A $(3,6)$-regular block code protograph and its corresponding base-matrix representation. The ``equal" circles \textcircled{=} represent variable nodes, while the ``plus" circles \textcircled{+} represent check nodes.}
    \label{BC_tanner}
\end{figure}
A block code protograph with design rate $R=b/c$ is a small bipartite graph with $c$ variable nodes and $c-b$ check nodes, which can be used to derive the graph of design rate $R=b/c$ block codes of various block sizes with the same degree distribution.\footnote{The term ``design rate" is used since the resulting parity-check matrix may have redundant rows. In this case, the code rate is slightly higher than the design rate.} An example of a block code protograph with $c=2$ variable nodes of degree 3 and $c-b=1$ check node of degree 6 is shown in Fig.~\ref{BC_tanner}. Let GF($q$) be a finite field with $q= 2^m$ elements, where $m$ is the number of bits used to represent a symbol over GF($q$). Let $M$~(typically a large integer) be the protograph {\em lifting factor}. A $q$-ary LDPC-BC with code length $n_{\rm BC}=Mc$ can be obtained from the $(c-b)\times c$ bi-adjacency matrix $\mathbf{B}=[B_{i,j}]$ of the protograph, called the {\em base matrix}, via the following two steps:
\begin{enumerate}
    \item   replace each nonzero entry $B_{i,j}$ in $\mathbf{B}$ with a summation of $B_{i,j}$ nonoverlapping $M \times M$ permutation matrices and each zero entry in $\mathbf{B}$ with the $M \times M$ all-zero matrix, where the elements $B_{i,j}$ in $\mathbf{B}$ are non-negative integers and the permutation matrices are chosen randomly and independently, resulting in a binary parity-check matrix $\mathbf{H}$ that is $M$ times as large as $\mathbf{B}$, and
    \item   replace the nonzero entries in $\mathbf{H}$ with randomly selected nonzero elements from the finite field GF($q$), resulting in a $q$-ary parity-check matrix $\mathbf{H}_\text{BC}$ of a $q$-ary LDPC-BC.
\end{enumerate}

For LDPC-BCs, data is typically transmitted in a sequence of independent blocks. At the decoder, an entire block must be received before BP decoding begins. Consequently, the decoding latency for a $q$-ary LDPC-BC constructed as described above over GF($q$), in terms of bits, is given by
\begin{equation}\label{BC_Latency}
    T_{\rm BC}=n_{\rm BC}\cdot m=Mmc.
\end{equation}

\subsection{SC-LDPC Codes over GF($q$)}\label{subsec:SC-LDPC}

\begin{figure}
    \center
    \includegraphics[angle=270, clip, width=\figwidth]{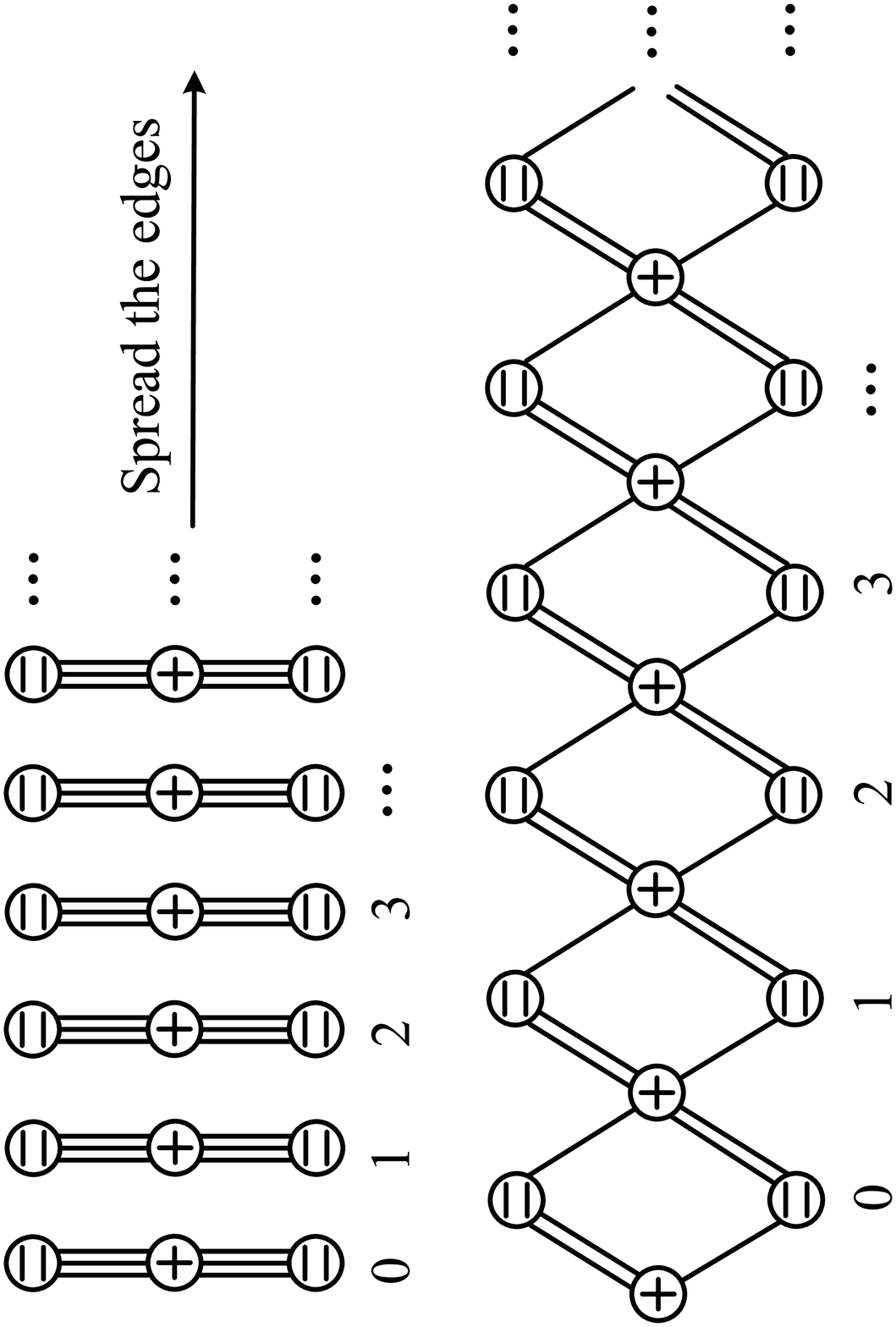}
    \caption{A $(3,6)$-regular SC-LDPC code protograph with $m_s=1$ constructed using the edge spreading procedure. The component submatrices used in the edge spreading are $\mathbf{B}_0=[2~1]$ and $\mathbf{B}_1=[1~2]$, where $\mathbf{B}=[3~3]$ is the base matrix of the underlying LDPC-BC.}
    \label{LDPC_tanner}
\end{figure}

Analogous to LDPC-BCs, SC-LDPC codes can also be derived using the protograph expansion method. Consider a $(c-b) \times c$ base matrix $\mathbf{B}$. We can use an edge spreading technique~\cite{lentmaier09} to construct a rate $R=b/c$ spatially coupled convolutional base matrix with syndrome former memory $m_s$ from $\mathbf{B}$ as
\begin{eqnarray}\label{submatrices}
\mathbf{B}_\text{SC}=
    \left[
    \begin{array}{ccc}
        \mathbf{B}_0 & & \\
        \mathbf{B}_1 &\mathbf{B}_0 &\\
        \vdots &\mathbf{B}_1 &\ddots\\
        \mathbf{B}_{m_s} &\vdots &\ddots\\
         &\mathbf{B}_{m_s} &\ddots\\
          & &\ddots
    \end{array}
    \right],
\end{eqnarray}
where the $m_s+1$ component submatrices $\mathbf{B}_0,\mathbf{B}_1,\ldots,\mathbf{B}_{m_s}$, each of size $(c-b) \times c$, satisfy
\begin{equation}\label{spread}
    \sum\limits_{i=0}^{m_s}\mathbf{B}_i =\mathbf{B}.
\end{equation}
An example of a rate $R=1/2$ $(3,6)$-regular SC-LDPC code protograph with $m_s=1$ constructed using the edge spreading procedure is shown in Fig.~\ref{LDPC_tanner}. The graph lifting operation is then applied to $\mathbf{B}_\text{SC}$ by replacing each nonzero entry in $\mathbf{B}_\text{SC}$ with~(a sum of) randomly selected permutation matrices of size $M \times M$ and each zero entry in $\mathbf{B}_\text{SC}$ with the $M \times M$ all-zero matrix, as described above, and then replacing the nonzero entries in the resulting convolutional parity-check matrix $\mathbf{H}_\text{SC}$ with randomly selected nonzero elements from the finite field GF($q$), resulting in an unterminated $q$-ary SC-LDPC code with constraint length $v_s = (m_s + 1)Mc$.\footnote{The constraint length determines the maximal width (in symbols) of the nonzero area of $\mathbf{H}_\text{SC}$.} The resulting $q$-ary SC-LDPC parity-check matrix $\mathbf{H}_\text{SC}$ is given
\ifCLASSOPTIONonecolumn
in~(\ref{eq:parity_check_matrix}), where the blank spaces in $\mathbf{H}_\text{SC}$ correspond to zeros and the submatrices $\mathbf{H}_i(t)$ have size $(c-b)M \times cM$, $\forall i,t$:
\fi
\ifCLASSOPTIONtwocolumn
in~(\ref{eq:parity_check_matrix}) at the top of the next page, where the blank spaces in $\mathbf{H}_\text{SC}$ correspond to zeros and the submatrices $\mathbf{H}_i(t)$ have size $(c-b)M \times cM$, $\forall i,t$.
\fi

\ifCLASSOPTIONtwocolumn
\begin{figure*}
\fi
\begin{equation}\label{eq:parity_check_matrix}
\mathbf{H}_\text{SC}=
\left[
\begin{array}{cccccc}
\mathbf{H}_0(0)       & & & &   &\\
\mathbf{H}_1(1)       &\mathbf{H}_0(1) & & &   &\\
\vdots                &\vdots                  &\ddots & &   &\\
\mathbf{H}_{m_s}(m_s) &\mathbf{H}_{m_s-1}(m_s) &\cdots                    &\mathbf{H}_{0}(m_s) &   &\\
                      &\mathbf{H}_{m_s}(m_s+1) &\mathbf{H}_{m_s-1}(m_s+1) &\cdots              &\mathbf{H}_{0}(m_s+1)  &\\
                      &                        &\ddots                    &                    &\ddots                 &\ddots
\end{array}
\right].
\end{equation}
\ifCLASSOPTIONtwocolumn
\hrulefill 
\end{figure*}
\fi

In this paper, we restrict consideration to SC-LDPC codes with syndrome former memory $m_s=1$, due to their superior performance with sliding window decoding~(see, e.g.,~\cite{iyengar2012windowed,Wei13,Wei13IT,lentmaier12,lentmaier11}). We also focus our attention on $(d_v,d_c)$-regular SC-LDPC codes, i.e., codes whose parity-check matrices have constant weight $d_v$ in each column and constant weight $d_c$ in each row, due to their complexity advantage compared to irregular codes and the fact that $(d_v,d_c)$-regular SC-LDPC code ensembles are capable of achieving capacity~(see~\cite{lentmaier10,kudekar11,kudekar13}).

In order to compare LDPC-BCs and SC-LDPC codes fairly, the freedom to select permutation matrices has been fixed in the following way. Consider two matrices $\mathbf{B}_0$ and $\mathbf{B}_1$, each of size $(c-b) \times c$, chosen such that $\mathbf{B}_0+\mathbf{B}_1$ is $(d_v,d_c)$-regular. The base matrix of a $(d_v,d_c)$-regular LDPC-BC is constructed as
\begin{eqnarray}\label{matrix_BC}
\mathbf{B}_\text{BC}=
    \left[\begin{array}{cccc}
        \mathbf{B}_0 &\mathbf{B}_1\\
        \mathbf{B}_1 &\mathbf{B}_0
        \end{array}
    \right]_{2(c-b) \times 2c},
\end{eqnarray}
where $\mathbf{B}_\text{BC}$ has weight $d_v$ in each column and weight $d_c$ in each row.\footnote{The ``weight" of a row~(column) of $\mathbf{B}_\text{BC}$ is the real sum of all the non-zero entries in the row~(column).} Then the block protograph expansion method described in Section~\ref{subsec:LDPC-BC} is used to form the parity-check matrix of a $(d_v,d_c)$-regular LDPC-BC as
\begin{eqnarray}\label{H_matrix_BC}
\mathbf{H}_\text{BC}=
    \left[\begin{array}{cccc}
        \mathbf{H}_0(0) &\mathbf{H}_1(2)\\
        \mathbf{H}_1(1) &\mathbf{H}_0(1)
        \end{array}
    \right]_{2(c-b)M \times 2cM}.
\end{eqnarray}
We construct the related SC-LDPC code in the following way. A $(d_v,d_c)$-regular SC-LDPC base matrix is constructed in the form of~(\ref{submatrices}) using component submatrices $\mathbf{B}_0$ and $\mathbf{B}_1$ as
\begin{eqnarray}\label{SC_B}
\mathbf{B}_\text{SC}=
    \left[
    \begin{array}{ccccc}
        \mathbf{B}_0 & & & &\\
        \mathbf{B}_1 &\mathbf{B}_0 & & &\\
                     &\mathbf{B}_1 &\mathbf{B}_0 & &\\
                     &             &\mathbf{B}_1 &\mathbf{B}_0 &\\
                     &             &             &\mathbf{B}_1 &\ddots\\
                     &             &             &             &\ddots
    \end{array}
    \right],
\end{eqnarray}
and a $(d_v,d_c)$-regular SC-LDPC parity-check matrix is then constructed using the usual protograph expansion method as
\begin{eqnarray}\label{H_matrix_SC}
\mathbf{H}_\text{SC}=
    \left[
    \begin{array}{ccccc}
        \mathbf{H}_0(0) & & & &\\
        \mathbf{H}_1(1) &\mathbf{H}_0(1) & & &\\
                     &\mathbf{H}_1(2) &\mathbf{H}_0(0) & &\\
                     &             &\mathbf{H}_1(1) &\mathbf{H}_0(1) &\\
                     &             &             &\mathbf{H}_1(2) &\ddots\\
                     &             &             &             &\ddots
    \end{array}
    \right].
\end{eqnarray}

\textbf{Remarks:} Note that the SC-LDPC code is time-varying with period 2, and its parity-check matrix $\mathbf{H}_\text{SC}$ uses exactly the same permutation matrices and elements from GF($q$) as $\mathbf{H}_\text{BC}$, now repeated periodically. This construction can be viewed as the unwrapping approach first presented in~\cite{jimenez99} for deriving an SC-LDPC code from an LDPC-BC. Note also that, even though we refer to a $(d_v,d_c)$-regular SC-LDPC base matrix and code, $\mathbf{B}_\text{SC}$ is not exactly $(d_v,d_c)$-regular, since its first~$(c-b)$ rows have weight less than $d_c$. This slight ``structured irregularity" associated with $(d_v,d_c)$-regular SC-LDPC codes is in fact the reason behind their capacity-approaching thresholds~(see, e.g.,~\cite{lentmaier10}).

\begin{table*}
\caption{Component matrices used in the construction of $(d_v,d_c)$-regular $q$-ary LDPC-BCs and $q$-ary SC-LDPC codes with field size $q=2^m$}\label{table0}
  \centering
  \begin{tabular}{|c||c|c|}
  \hline
  Codes &Component matrices &Block/constraint length\\\hline
  $(2,4)$-regular &$\mathbf{B}_{0}=\mathbf{B}_{1}=[1~1]$ &$4Mm$\\\hline
  $(3,6)$-regular &$\mathbf{B}_{0}=[2~1]$, $\mathbf{B}_{1}=[1~2]$ &$4Mm$\\\hline
  $(3,9)$-regular &$\mathbf{B}_{0}=[1~2~2]$, $\mathbf{B}_{1}=[2~1~1]$ &$6Mm$\\\hline
  $(3,12)$-regular &$\mathbf{B}_{0}=[1~1~2~2]$, $\mathbf{B}_{1}=[2~2~1~1]$ &$8Mm$\\\hline
\end{tabular}
\end{table*}
The parity-check matrices $\mathbf{H}_\text{BC}$ and $\mathbf{H}_\text{SC}$ of $(d_v,d_c)$-regular $q$-ary LDPC-BCs and $q$-ary SC-LDPC codes are constructed over GF($q$) in the form of~(\ref{H_matrix_BC}) and~(\ref{H_matrix_SC}), respectively, using the component submatrices shown in Table~\ref{table0}. Given a protograph lifting factor $M$, the block length~(in bits) of the $(d_v,d_c)$-regular $q$-ary LDPC-BCs and the constraint length~(in bits) of the $(d_v,d_c)$-regular $q$-ary SC-LDPC codes are both equal to $2Mmc$, where the field size is $q=2^m$.

\section{Pipeline and Sliding Window Decoding for SC-LDPC Codes over GF($q$)}\label{sec:Decoding}
Although the Tanner graph of a $q$-ary SC-LDPC code has an infinite number of nodes, the distance between two variable nodes that are connected to the same check node is limited by the constraint length of the code. This restriction gives rise to efficient decoder implementations such as the high-throughput pipeline decoder~\cite{jimenez99,Ali08} and the low-latency sliding window decoder~\cite{lentmaier10,iyengar2012windowed,lentmaier11}.

\subsection{Pipeline Decoding}
\begin{figure*}
    \center
    \includegraphics[angle=270, clip, width=\Twofigwidth]{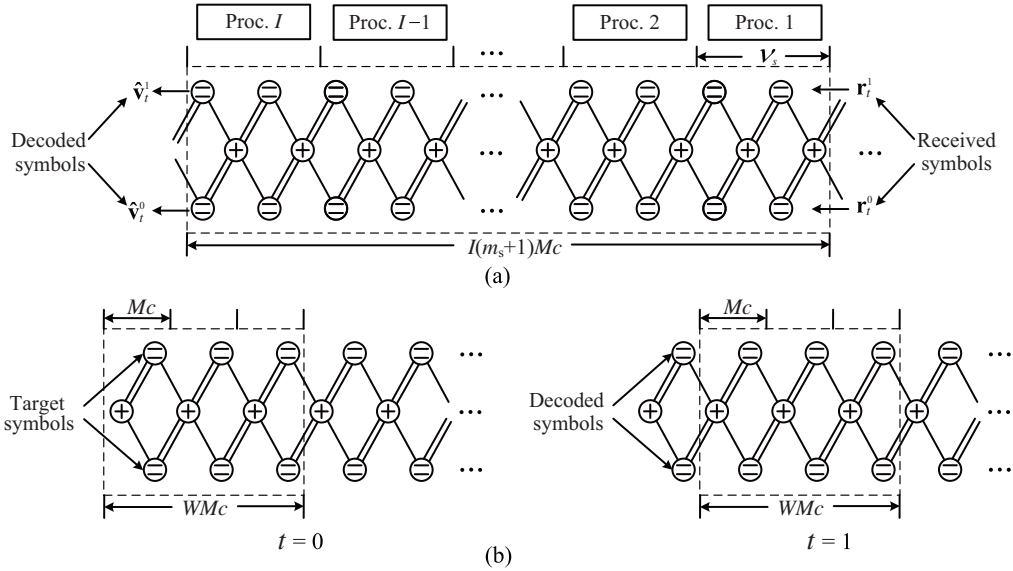}
    \caption{(a)~Example of a pipeline decoder operating on the protograph of a $(3,6)$-regular $q$-ary SC-LDPC code with $m_s=1$. (b)~Example of a sliding window decoder with window size $W=3$ operating on the protograph of the same $(3,6)$-regular $q$-ary SC-LDPC code with $m_s=1$ at times $t=0$~(left), and $t=1$~(right).}
    \label{sliding_window}
\end{figure*}

An example of a pipeline decoder operating on the protograph of a $(3,6)$-regular $q$-ary SC-LDPC code with $m_s=1$ is shown in Fig.~\ref{sliding_window}(a). Given some fixed number $I$ of decoding iterations, the pipeline decoder employs $I$ identical copies of a message-passing processor operating in parallel.\footnote{A serial decoding architecture~\cite{bates2008low} can be used to reduce the number of processors at a cost of reduced throughput.} Each processor includes only one constraint length, i.e., $v_s = (m_s + 1)Mc$, of variable nodes, and during a single decoding iteration messages are only passed within a single processor, so equating the processor complexity of SC-LDPC codes and LDPC-BCs means equating the constraint length of SC-LDPC codes to the block length of LDPC-BCs~\cite{Ali08,pusane2011deriving}. Note that $Iv_s = I(m_s + 1)Mc$ represents the total decoding latency in received symbols and the total number of soft received values that must be stored in the decoder memory at any given time. Since capacity approaching performance can require a large number of iterations~$I$, these latency and memory requirements of pipeline decoding may be unacceptably high.

\subsection{Sliding Window Decoding}
In this subsection, we propose a sliding window decoding architecture for $q$-ary SC-LDPC codes, which is an extension of the sliding window decoding architecture presented in~\cite{iyengar2012windowed} for binary SC-LDPC codes.

An example of a sliding window decoder with window size $W=3$ operating on the protograph of a $(3,6)$-regular $q$-ary SC-LDPC code with $m_s=1$ is shown in Fig.~\ref{sliding_window}(b). Assuming a window size of $WMc$ symbols, decoding proceeds until a fixed number of iterations has been performed or some stopping rule~(see~Section~\ref{subsec:StoppingRules}) is satisfied, after which the window shifts $Mc$ positions and the $Mc$ symbols shifted out of the window are decoded. The first $Mc$ symbols in any window are called {\em target symbols}. The decoding latency of the sliding window decoder for $q$-ary SC-LDPC codes, in terms of bits, is given by
\begin{equation}\label{SC_Latency}
  T_{\rm SC}=WMmc.
\end{equation}
The iterative decoding algorithm within a window can be implemented with existing algorithms, such as the FFT-QSPA~\cite{Barnault03}, EMS algorithms~\cite{Declercq07,Voicila10,Ma12}, and so on.
\subsection{A Stopping Rule for Sliding Window Decoding}\label{subsec:StoppingRules}
For LDPC-BCs, iterative decoding is stopped if the decoded sequence is a valid codeword, i.e., if and only if all of the parity-check equations are satisfied. However, this stopping rule cannot be used with sliding window decoding of SC-LDPC codes, because we only decode one set of target symbols at a time. In this subsection, we propose a stopping rule based on a soft BER estimate for sliding window decoding of $q$-ary SC-LDPC codes, which is motivated by the method presented in~\cite{lentmaier12}.

Let $P_{t}^{(j)}(b)$ for $0\leq j <Mc$ be the probability that the $j$-th symbol $v_t^{(j)}$ in a window at time $t$ is $b \in$ GF($q$), given the decoder input from the channel and the constraints of the $q$-ary SC-LDPC code. After each iteration of the BP algorithm at time $t$, we make hard decisions $\hat{v}_t^{(j)}$ on $v_t^{(j)}$ based on the probabilities $P_{t}^{(j)}(x),x \in$ GF($q$), computed at the decoder by choosing $\hat{v}_t^{(j)}=x$ as the symbol with the maximum probability. The probability that $\hat{v}_t^{(j)}$ is wrong is then given by
\begin{equation}\label{error}
e_t^{(j)} = 1-P_{t}^{(j)}(x=\hat{v}_t^{(j)}),
\end{equation}
and the estimated soft BER $\hat{P}_t$ can be calculated as
\begin{equation}\label{targetedBER}
\hat{P}_t = \frac{1}{Mc}\sum\limits_{j=0}^{Mc-1}e_t^{(j)}.
\end{equation}
The proposed stopping rule is as follows: the window shifts only when either a fixed number of iterations $I_{\rm max}$ has been performed or $\hat{P}_t$ is less than a preselected target BER.

In the simulation results presented in this paper, the nodes within a decoding window are updated according to a uniform parallel~(flooding) schedule, so that all the nodes within the window are updated in parallel during each decoding iteration. Note, however, that the node updates can also be performed serially and/or non-uniformly in order to reduce computational complexity~(see, e.g.,~\cite{Hassan12,Hassan13}).

\section{An Equal Block Length and Constraint Length Comparison}\label{sec:CG}
In this section, we focus on the case of equal decoder processor~(hardware) complexity, i.e., when the constraint length of the $q$-ary SC-LDPC codes is equal to the block length of the $q$-ary LDPC-BCs.\footnote{It should be noted that, in this case, the latency of the SC-LDPC code is higher than for the LDPC-BC. An equal latency comparison is the subject of the next section.} We consider binary phase-shift keying~(BPSK) modulation over the BI-AWGN channel. For $q$-ary LDPC-BCs, the FFT-QSPA with the parity-check-based stopping rule is applied with $I_{\rm max}$ set to 100. For $q$-ary SC-LDPC codes, sliding window decoding is also implemented with the FFT-QSPA, $I_{\rm max}$ is set to 100, and the stopping rule proposed in Section~\ref{subsec:StoppingRules} with a preselected target BER of $10^{-6}$.

\subsection{$(2,4)$-Regular LDPC Codes over GF($q$)}

\begin{figure}
    \center
    \includegraphics[clip, width=\figwidth]{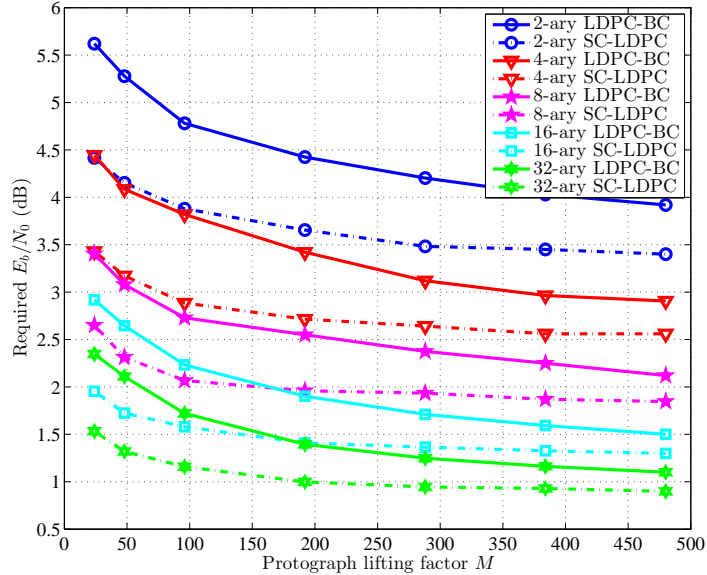}
    \caption{Required $E_b/N_0$ to achieve a BER of $10^{-4}$ with different protograph lifting factors $M$ for $(2,4)$-regular codes over GF($2$), GF($4$), GF($8$), GF($16$), and GF($32$). The window size of the sliding window decoder is $W=12$. Solid curves represent LDPC-BCs, while dotted curves represent SC-LDPC codes.}
    \label{Fig_2_4}
\end{figure}
The values of the bit signal-to-noise ratio~(SNR) $E_b/N_0$ needed to achieve a BER of $10^{-4}$ with different protograph lifting factors $M$ for rate $R=1/2$ $(2,4)$-regular codes over GF($2$), GF($4$), GF($8$), GF($16$), and GF($32$) are shown in Fig.~\ref{Fig_2_4}.\footnote{We choose BERs of $10^{-4}$~($10^{-5}$ in Section~\ref{sec:ELC}) for comparison because they represent target BERs commonly used in many practical applications.} The window size of the sliding window decoder for the $q$-ary SC-LDPC codes is $W=12$. From Fig.~\ref{Fig_2_4}, we see that the performance of $(2,4)$-regular $q$-ary LDPC-BCs and $q$-ary SC-LDPC codes improves as the protograph lifting factor $M$ increases. We also see that $(2,4)$-regular $q$-ary SC-LDPC codes with short constraint length~(corresponding to small $M$) achieve substantial ``convolutional gains" compared to the underlying LDPC-BCs, but the gains diminish as the protograph lifting factor $M$ increases. For example, the convolutional gain of the SC-LDPC code compared to the LDPC-BC over GF($16$) when $M=24$ is about $1.0$ dB, but it decreases to only $0.2$ dB when $M=480$. These results are consistent with the asymptotic~(large $M$) threshold performance analysis presented in~\cite{Wei13IT}, where the thresholds of $(2,4)$-regular SC-LDPC codes with these field sizes are shown to be only slightly better than those of $(2,4)$-regular LDPC-BCs.

It is also observed in~\cite{Wei13IT} that, compared to $(2,4)$-regular $q$-ary SC-LDPC codes, $(d_v,d_c)$-regular $q$-ary SC-LDPC codes with $d_v\geq 3$ provide capacity-approaching performance using window decoding when both the field size $q$ and the window size $W$ are relatively small. Since small $q$ is desirable to reduce complexity and small $W$ is desirable to reduce latency, we focus on $(d_v,d_c)$-regular $q$-ary LDPC codes with $d_v\geq 3$ in the rest of the paper.

\subsection{$(3,6)$-Regular LDPC Codes over GF($q$)}\label{sec:CG_36}

\begin{figure}
    \center
    \includegraphics[clip, width=\figwidth]{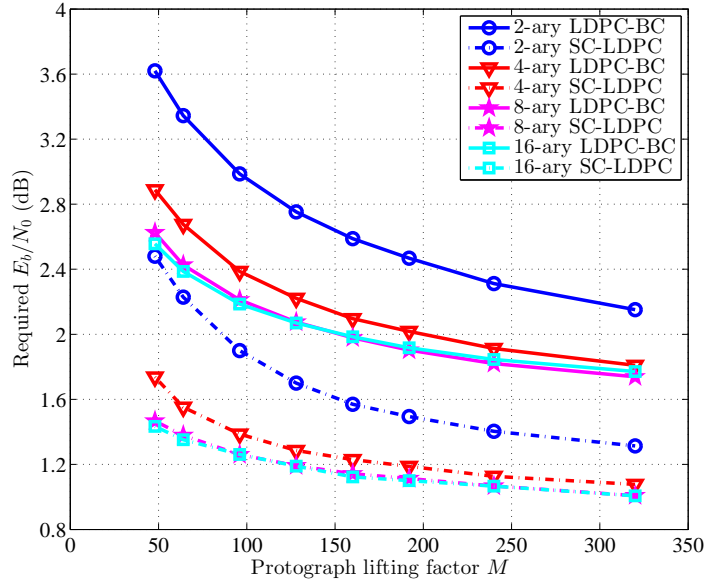}
    \caption{Required $E_b/N_0$ to achieve a BER of $10^{-4}$ with different protograph lifting factors $M$ for $(3,6)$-regular codes over GF($2$), GF($4$), GF($8$), and GF($16$). The window size of the sliding window decoder is $W=12$. Solid curves represent LDPC-BCs, while dotted curves represent SC-LDPC codes.}
    \label{Fig_3_6}
\end{figure}
The values of $E_b/N_0$ needed to achieve a BER of $10^{-4}$ with different protograph lifting factors $M$ for rate $R=1/2$ $(3,6)$-regular codes over GF($2$), GF($4$), GF($8$), and GF($16$) are shown in Fig.~\ref{Fig_3_6}. The window size of the sliding window decoder for the $q$-ary SC-LDPC codes is $W=12$. Similar to the $(2,4)$-regular $q$-ary codes, we see in Fig.~\ref{Fig_3_6} that the performance of the $(3,6)$-regular $q$-ary LDPC-BCs and $q$-ary SC-LDPC codes improves as the protograph lifting factor $M$ increases. We also observe that $(3,6)$-regular $q$-ary SC-LDPC codes achieve substantial convolutional gains compared to the underlying LDPC-BCs over the entire range of lifting factors, with the amount of gain declining gradually as $M$ increases. For example, the convolutional gain of the SC-LDPC code compared to the LDPC-BC over GF($8$) when $M=48$ is about $1.1$ dB, and it decreases to around $0.8$ dB for $M=320$. By comparing Figs.~\ref{Fig_2_4} and~\ref{Fig_3_6}, we see that the convolutional gains, relative to the LDPC-BCs, of the $(3,6)$-regular SC-LDPC codes are larger than those of the $(2,4)$-regular SC-LDPC codes. This is again consistent with the asymptotic threshold performance analysis presented in~\cite{Wei13IT}, where the thresholds of $(3,6)$-regular SC-LDPC codes are shown to be substantially better than those of $(3,6)$-regular LDPC-BCs.

\textbf{Remark:} Although it has been reported in~\cite{lentmaier10} that the BP thresholds of $(4,8)$-regular binary SC-LDPC codes are better than those of $(3,6)$-regular binary SC-LDPC codes, we found from simulation that $(3,6)$-regular $q$-ary SC-LDPC codes perform better than $(4,8)$-regular $q$-ary SC-LDPC codes at~(low) SNRs and when~(short-to-moderate) constraint lengths are considered, i.e., $(4,8)$-regular SC-LDPC codes typically require a large lifting factor $M$ to outperform $(3,6)$-regular SC-LDPC codes. This is consistent with the discussion concerning the practical design of SC-LDPC codes in Section VI-A of~\cite{kudekar11}, where it is noted that large~(variable node and check node) degrees imply slower convergence for finite-length ensembles to the asymptotic performance limit. For these reasons, we focus the rest of our discussion on $(d_v,d_c)$-regular $q$-ary SC-LDPC codes for which the variable node degree is fixed at $d_v=3$.

\ifCLASSOPTIONonecolumn
\begin{figure*}[htbp]
\fi
\ifCLASSOPTIONtwocolumn
\begin{figure}[htbp]
\fi
\centering
\subfigure[$(3,9)$-regular $q$-ary LDPC codes]{
\label{Fig_3_9_CG}
\includegraphics[clip, width=\figwidth]{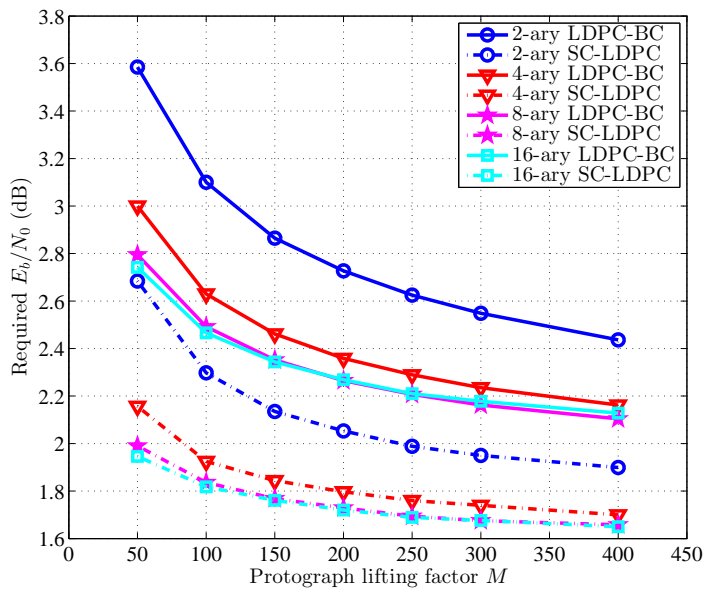}}
\subfigure[$(3,12)$-regular $q$-ary LDPC codes]{
\label{Fig_3_12_CG}
\includegraphics[clip, width=\figwidth]{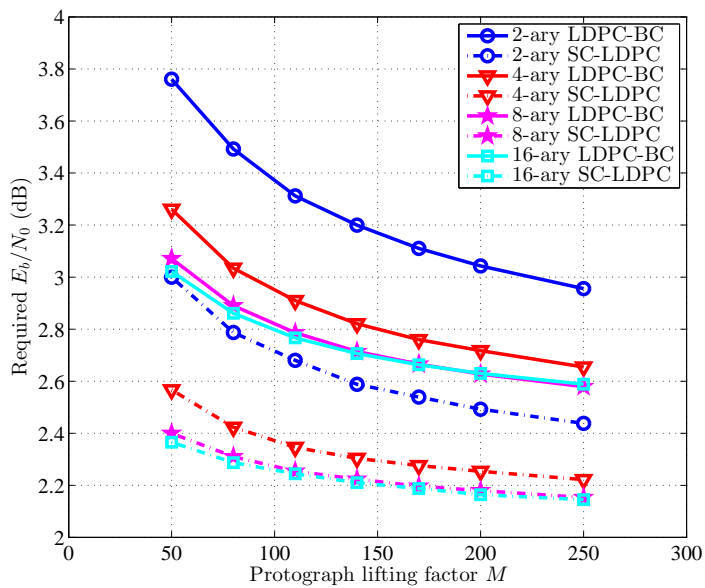}}
\caption{Required $E_b/N_0$ to achieve a BER of $10^{-4}$ with different protograph lifting factors $M$ for high-rate codes over GF($2$), GF($4$), GF($8$), and GF($16$). The window size of the sliding window decoder is $W=12$. Solid curves represent LDPC-BCs, while dotted curves represent SC-LDPC codes.}
\label{Fig_HighRate_CG}
\ifCLASSOPTIONonecolumn
\end{figure*}
\fi
\ifCLASSOPTIONtwocolumn
\end{figure}
\fi
\subsection{High-Rate LDPC Codes over GF($q$)}\label{sec:CG_3_9_12}
The values of $E_b/N_0$ needed to achieve a BER of $10^{-4}$ with different protograph lifting factors $M$ for rate $R=2/3$ and $3/4$ $(3,9)$- and $(3,12)$-regular codes over GF($2$), GF($4$), GF($8$), and GF($16$) are shown in Fig.~\ref{Fig_HighRate_CG}.  The window size of the sliding window decoder for the $q$-ary SC-LDPC codes is $W=12$. From Fig.~\ref{Fig_HighRate_CG}, we see that the performance of $(3,9)$-regular and $(3,12)$-regular $q$-ary LDPC-BCs and $q$-ary SC-LDPC codes improves as the protograph lifting factor $M$ increases. We also observe that both $(3,9)$-regular and $(3,12)$-regular $q$-ary SC-LDPC codes achieve substantial convolutional gains compared to the underlying LDPC-BCs over the entire range of lifting factors, with the amount of gain declining gradually as $M$ increases. This is again consistent with the asymptotic threshold performance analysis presented in~\cite{Wei13IT}, where the thresholds of $(3,9)$- and $(3,12)$-regular SC-LDPC codes are shown to be substantially better than those of $(3,9)$- and $(3,12)$-regular LDPC-BCs, respectively.

\section{An Equal Latency Comparison}\label{sec:ELC}
In addition to decoding performance, the latency introduced by employing channel coding is a crucial factor in the design of a practical communication system. For example, minimizing latency is of major importance in applications such as personal wireless communication, real-time audio and video, and command and control military communication. In this section, we consider the case when the decoding latency of $q$-ary SC-LDPC codes and $q$-ary LDPC-BCs is the same.

\subsection{$(3,6)$-Regular LDPC Codes over GF($q$)}\label{subsec:3_6}
For the rate $R=1/2$ $(3,6)$-regular $q$-ary SC-LDPC codes with $\mathbf{H}_\text{SC}$ given by~(\ref{H_matrix_SC}), the decoding latency of the sliding window decoder is given by
\begin{equation}\label{3_6_SC_latency}
    T_{\rm SC}=2WM_{\rm SC}m,
\end{equation}
whereas the rate $R=1/2$ $(3,6)$-regular $q$-ary LDPC-BCs with $\mathbf{H}_\text{BC}$ given by~(\ref{H_matrix_BC}) have decoding latency
\begin{equation}\label{3_6_BC_latency}
T_{\rm BC}=4M_{\rm BC}m,
\end{equation}
where we now distinguish between the lifting factors $M_{\rm SC}$ of the SC-LDPC codes and $M_{\rm BC}$ of the LDPC-BCs.

\begin{figure}
        \center
        \includegraphics[clip, width=\figwidth]{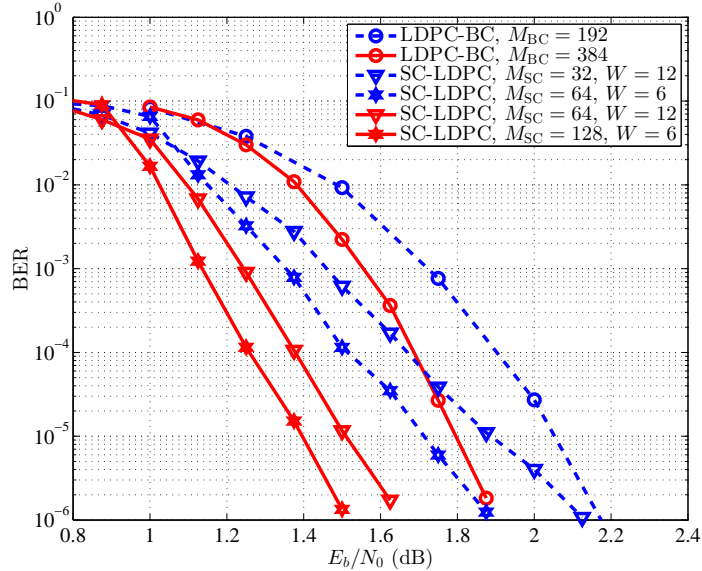}
        \caption{Simulated decoding performance of $(3,6)$-regular 8-ary SC-LDPC codes compared to $(3,6)$-regular 8-ary LDPC-BCs with protograph lifting factors $M_{\rm BC}=192$ and $M_{\rm BC}=384$. The values of $M_{\rm SC}$ and $W$ for the SC-LDPC codes with sliding window decoding are chosen in such a way that the decoding latency is equal to the block length of the LDPC-BC.}
        \label{Fig_SameLatency_3_6}
\end{figure}
In Fig.~\ref{Fig_SameLatency_3_6}, $(3,6)$-regular 8-ary SC-LDPC codes are compared to $(3,6)$-regular 8-ary LDPC-BCs and the values of the protograph lifting factors $M_{\rm SC}$ and $M_{\rm BC}$ are chosen such that the decoding latency of the LDPC-BCs and the SC-LDPC codes are the same. Even in this case, we see that the performance of the SC-LDPC codes is still significantly better than that of the LDPC-BCs. From Fig.~\ref{Fig_SameLatency_3_6}, we also see that the SC-LDPC code constructed with a larger lifting factor $M_{\rm SC}$ and decoded with a smaller window size $W=6$ outperforms the SC-LDPC code constructed with a smaller $M_{\rm SC}$ and decoded with a larger window size $W=12$~(both have the same decoding latency). In other words, selecting a smaller $W$, which is typically detrimental to decoder performance, is compensated for by allowing a larger $M_{\rm SC}$, which improves code performance. For example, at a BER of $10^{-5}$, the 8-ary SC-LDPC code with $M_{\rm SC}=64$ and decoded with window size $W=12$ gains $0.3$ dB compared to the equal latency 8-ary LDPC-BC with $M_{\rm BC}=384$, while the gain increases to $0.4$ dB by using the 8-ary SC-LDPC code with $M_{\rm SC}=128$ and $W=6$. Similar behavior for binary SC-LDPC codes was reported in~\cite{lentmaier12,lentmaier11}.

\begin{figure}
        \center
        \includegraphics[clip, width=\figwidth]{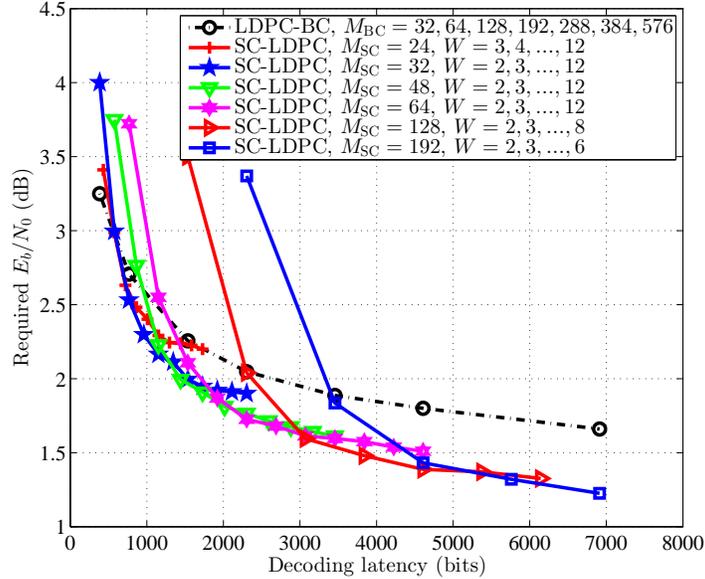}
        \caption{Required $E_b/N_0$ to achieve a BER of $10^{-5}$ for $(3,6)$-regular 8-ary LDPC-BCs and $(3,6)$-regular 8-ary SC-LDPC codes as a function of decoding latency.}
        \label{Fig_DiffLatency_3_6}
\end{figure}

The $E_b/N_0$ required to achieve a BER of $10^{-5}$ for equal latency $(3,6)$-regular $8$-ary LDPC-BCs and $(3,6)$-regular 8-ary SC-LDPC codes as a function of decoding latency is shown in Fig.~\ref{Fig_DiffLatency_3_6}, where we observe that the performance of the SC-LDPC codes~(with fixed protograph lifting factor $M_{\rm SC}$) improves as the window size $W$~(and hence the latency) increases, but it does not improve much further beyond a certain window size~(roughly $W=10$). Also, beyond a certain latency, using a larger protograph lifting factor $M_{\rm SC}$ with a smaller window size $W$ gives better performance. For example, when the decoding latency is $2304$ bits, the performance of the 8-ary SC-LDPC code with $M_{\rm SC}=64$ and decoded with $W=6$ is better than that of the SC-LDPC code with $M_{\rm SC}=32$ and decoded with $W=12$ and, when the decoding latency is $4608$ bits, the performance with $M_{\rm SC}=128$ and $W=6$ is better than with $M_{\rm SC}=64$ and $W=12$. Furthermore, we observe that the LDPC-BCs always perform worse than the SC-LDPC codes except when either $M_{\rm SC}$ and/or $W$ are too small.

Note that increasing the window size $W$ improves decoder performance and increasing the protograph lifting factor $M_{\rm SC}$ improves code performance. For example, from Fig.~\ref{Fig_DiffLatency_3_6} we see that when the decoding latency is 2304 bits, the decoding performance of the 8-ary SC-LDPC code with $M_{\rm SC}=64$ and decoded with $W=6$ is better than that of the SC-LDPC code with $M_{\rm SC}=128$ and decoded with $W=3$, the reverse of the situation for the same codes when the latency is 4608 bits~(obtained for window sizes $W=12$ and $W=6$, respectively). In this case, for a latency of 2304 bits, the performance loss caused by the small window size~($W=3$) is not compensated for by the larger lifting factor~($M_{\rm SC}=128$), whereas, if we double the window sizes~(increasing the latency to 4608 bits), the code with the larger lifting factor~($M_{\rm SC}=128$) has a large enough window size~($W=6$) to outperform the smaller lifting factor~($M_{\rm SC}=64$) code with $W=12$. This raises the interesting question of how to choose $M_{\rm SC}$ and $W$ in order to achieve the best performance when the decoding latency of the sliding window decoder is fixed.

\begin{figure}
        \center
        \includegraphics[clip, width=\figwidth]{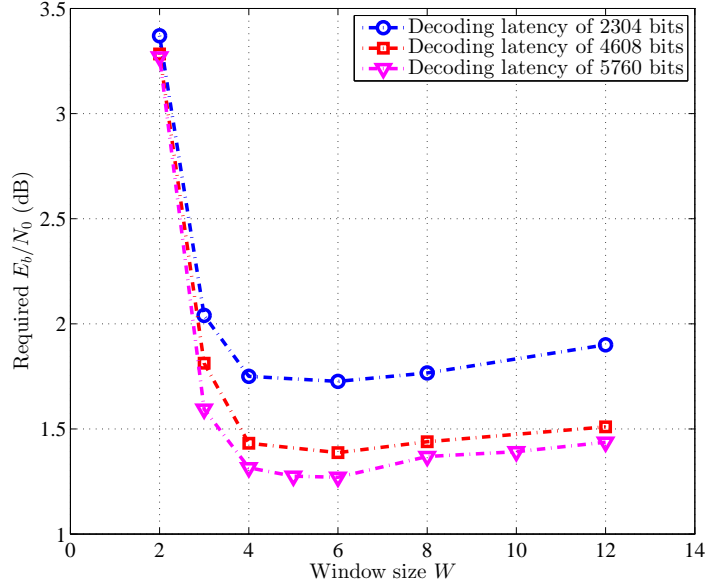}
        \caption{Required $E_b/N_0$ to achieve a BER of $10^{-5}$ for $(3,6)$-regular 8-ary SC-LDPC codes with different window sizes $W$ and decoding latencies of 2304, 4608, and 5760 bits.}
        \label{Fig_SameLatency_VS}
\end{figure}
\begin{figure}
    \center
    \includegraphics[clip, width=\figwidth]{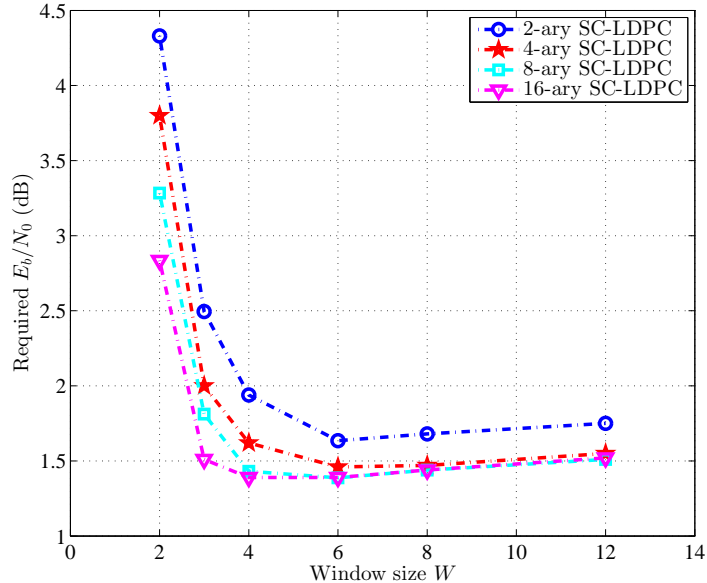}
    \caption{Required $E_b/N_0$ to achieve a BER of $10^{-5}$ for $(3,6)$-regular $q$-ary SC-LDPC codes with different window sizes $W$ when the decoding latency is 4608 bits.}
    \label{Fig_4608bits_diff_Fileds}
\end{figure}

Fig.~\ref{Fig_SameLatency_VS} shows the $E_b/N_0$ required for $(3,6)$-regular 8-ary SC-LDPC codes to achieve a BER of $10^{-5}$ with different window sizes $W$ and decoding latencies of 2304, 4608, and 5760 bits. We observe that the required $E_b/N_0$ decreases dramatically until around $W=4$ to $W=6$, and then it increases gradually as the window size $W$ increases. This increase results from the fact that the improved decoder performance obtained by increasing $W$ is not compensating for the decrease in code performance as a result of the smaller lifting factor. We therefore conclude that, for $(3,6)$-regular 8-ary SC-LDPC codes, $W=6$ is a good choice for optimum performance. Similar behavior has also been observed for other field sizes, as shown in Fig.~\ref{Fig_4608bits_diff_Fileds}.

\begin{table*}
\caption{Minimum $E_b/N_0$ required to achieve a BER of $10^{-5}$ for $(3,6)$-regular $q$-ary LDPC-BCs and $(3,6)$-regular $q$-ary SC-LDPC codes with different field sizes and decoding latencies of 2304, 4608, 6912, 9216, and 13824 bits}\label{table1}
  \centering
  \begin{tabular}{|c||c|c|c|c|c|c|c|c|}
  \hline
  \multirow{2}{*}{Required $\rm{E_b/N_0}$ (dB)} &\multicolumn{4}{c|}{LDPC-BC} &\multicolumn{4}{c|}{SC-LDPC~($W=6$)}\\ \cline{2-9}
  &GF($2$) &GF($4$) &GF($8$) &GF($16$) &GF($2$) &GF($4$) &GF($8$) &GF($16$) \\ \hline
  Latency of 2304 bits &2.1 &2.0 &2.0 &2.2 &2.3 &1.9 &1.7 &1.7 \\\hline
  Latency of 4608 bits &1.8 &1.7 &1.8 &1.9 &1.6 &1.5 &1.4 &1.4 \\\hline
  Latency of 6912 bits &1.7 &1.6 &1.7 &1.8 &1.5 &1.3 &1.2 &1.2 \\\hline
  Latency of 9216 bits &1.6 &1.5 &1.6 &1.7 &1.3 &1.2 &1.1 &1.1 \\\hline
  Latency of 13824 bits &1.5 &1.4 &1.5 &1.6 &1.2 &1.1 &1.0 &1.0 \\\hline
\end{tabular}
\end{table*}
Table~\ref{table1} shows the minimum $E_b/N_0$ required to achieve a BER of $10^{-5}$ for some $(3,6)$-regular $q$-ary LDPC-BCs and $(3,6)$-regular $q$-ary SC-LDPC codes with different field sizes and decoding latencies of 2304, 4608, 6912, 9216, and 13824 bits. It is observed that the non-binary SC-LDPC codes outperform both the binary and non-binary LDPC-BCs and the binary SC-LDPC codes for fixed decoding latency. In general, in contrast to $q$-ary LDPC-BCs, the required $E_b/N_0$ for $q$-ary SC-LDPC codes to achieve a BER of $10^{-5}$ decreases as we increase the field size $q$. This is consistent with results obtained for the iterative decoding thresholds in~\cite{Wei13IT}, where it is shown that, for increasing $q$, the thresholds of $(3,6)$-regular $q$-ary SC-LDPC codes approach capacity, but those of $(3,6)$-regular $q$-ary LDPC-BCs diverge from capacity. Finally, note that, for a latency of 2304 bits, the minimum $E_b/N_0$ required to achieve a BER of $10^{-5}$ for $(3,6)$-regular binary SC-LDPC codes is higher than for $(3,6)$-regular binary LDPC-BCs, which is due to the error floor effect of binary SC-LDPC codes with short constraint lengths. This effect is not observed at higher BERs or larger latencies, as can be seen for latencies of 4608, 6912, 9216, and 13824 bits, where binary SC-LDPC codes outperform binary LDPC-BCs.

\subsection{High-Rate LDPC Codes over GF($q$)}
For rate $R=2/3$ $(3,9)$-regular $q$-ary SC-LDPC codes, the decoding latency of the sliding window decoder is given by
\begin{equation}\label{3_9_SC_latency}
    T_{\rm SC}=3WM_{\rm SC}m,
\end{equation}
whereas $R=2/3$ $(3,9)$-regular $q$-ary LDPC-BCs have decoding latency
\begin{equation}\label{3_9_BC_latency}
T_{\rm BC}=6M_{\rm BC}m.
\end{equation}
For $R=3/4$ $(3,12)$-regular $q$-ary SC-LDPC codes, the decoding latency of the sliding window decoder is given by
\begin{equation}\label{3_12_SC_latency}
    T_{\rm SC}=4WM_{\rm SC}m,
\end{equation}
whereas $R=3/4$ $(3,12)$-regular $q$-ary LDPC-BCs have decoding latency
\begin{equation}\label{3_12_BC_latency}
T_{\rm BC}=8M_{\rm BC}m.
\end{equation}

\ifCLASSOPTIONonecolumn
\begin{figure*}[htbp]
\fi
\ifCLASSOPTIONtwocolumn
\begin{figure}[htbp]
\fi
\centering
\subfigure[$(3,9)$-regular 8-ary LDPC codes]{
\label{Fig_3_9_EL}
\includegraphics[clip, width=\figwidth]{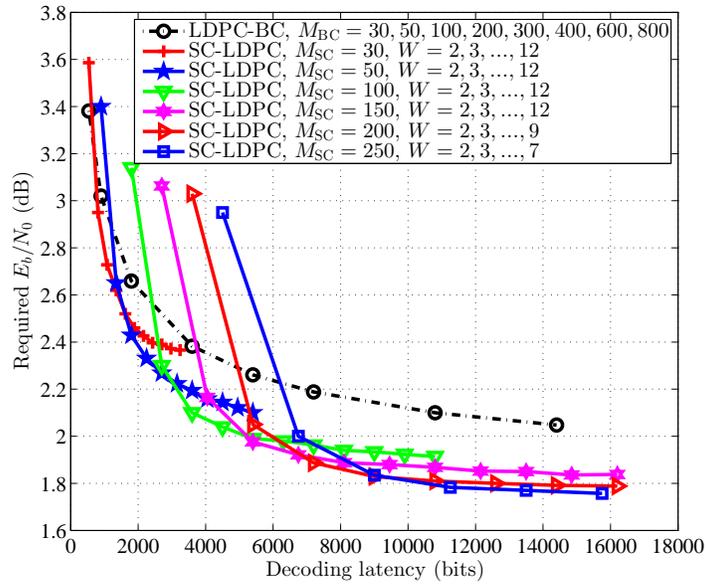}}
\subfigure[$(3,12)$-regular 8-ary LDPC codes]{
\label{Fig_3_12_EL}
\includegraphics[clip, width=\figwidth]{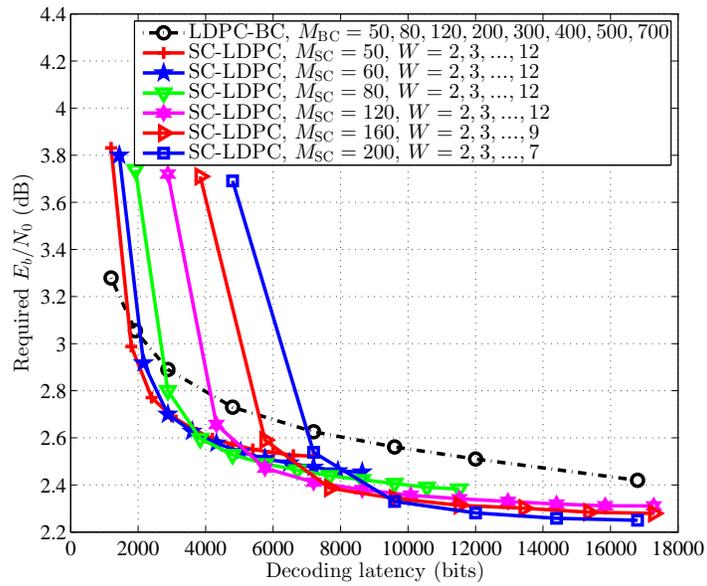}}
\caption{Required $E_b/N_0$ to achieve a BER of $10^{-5}$ for high-rate 8-ary LDPC-BCs and 8-ary SC-LDPC codes as a function of decoding latency.}
\label{Fig_HighRate_EqualLatency}
\ifCLASSOPTIONonecolumn
\end{figure*}
\fi
\ifCLASSOPTIONtwocolumn
\end{figure}
\fi

The $E_b/N_0$ required to achieve a BER of $10^{-5}$ for equal latency $(3,9)$-regular and $(3,12)$-regular $8$-ary LDPC-BCs and SC-LDPC codes as a function of decoding latency is shown in Fig.~\ref{Fig_HighRate_EqualLatency}. Similar to the $(3,6)$-regular 8-ary case, we observe that the performance of both $(3,9)$- and $(3,12)$-regular SC-LDPC codes~(with fixed protograph lifting factor $M_{\rm SC}$) improves as the window size $W$ increases, but it does not improve much beyond a certain window size~(roughly $W=8$). Moreover, under an equal latency constraint, both $(3,9)$- and $(3,12)$-regular LDPC-BCs always perform worse than the corresponding $(3,9)$- and $(3,12)$-regular SC-LDPC codes except when either $M_{\rm SC}$ and/or $W$ are too small.

\ifCLASSOPTIONonecolumn
\begin{figure*}[htbp]
\fi
\ifCLASSOPTIONtwocolumn
\begin{figure}[htbp]
\fi
\centering
\subfigure[$(3,9)$-regular 8-ary SC-LDPC codes]
{
\label{Fig_3_9_ELGF8}
\includegraphics[clip, width=\figwidth]{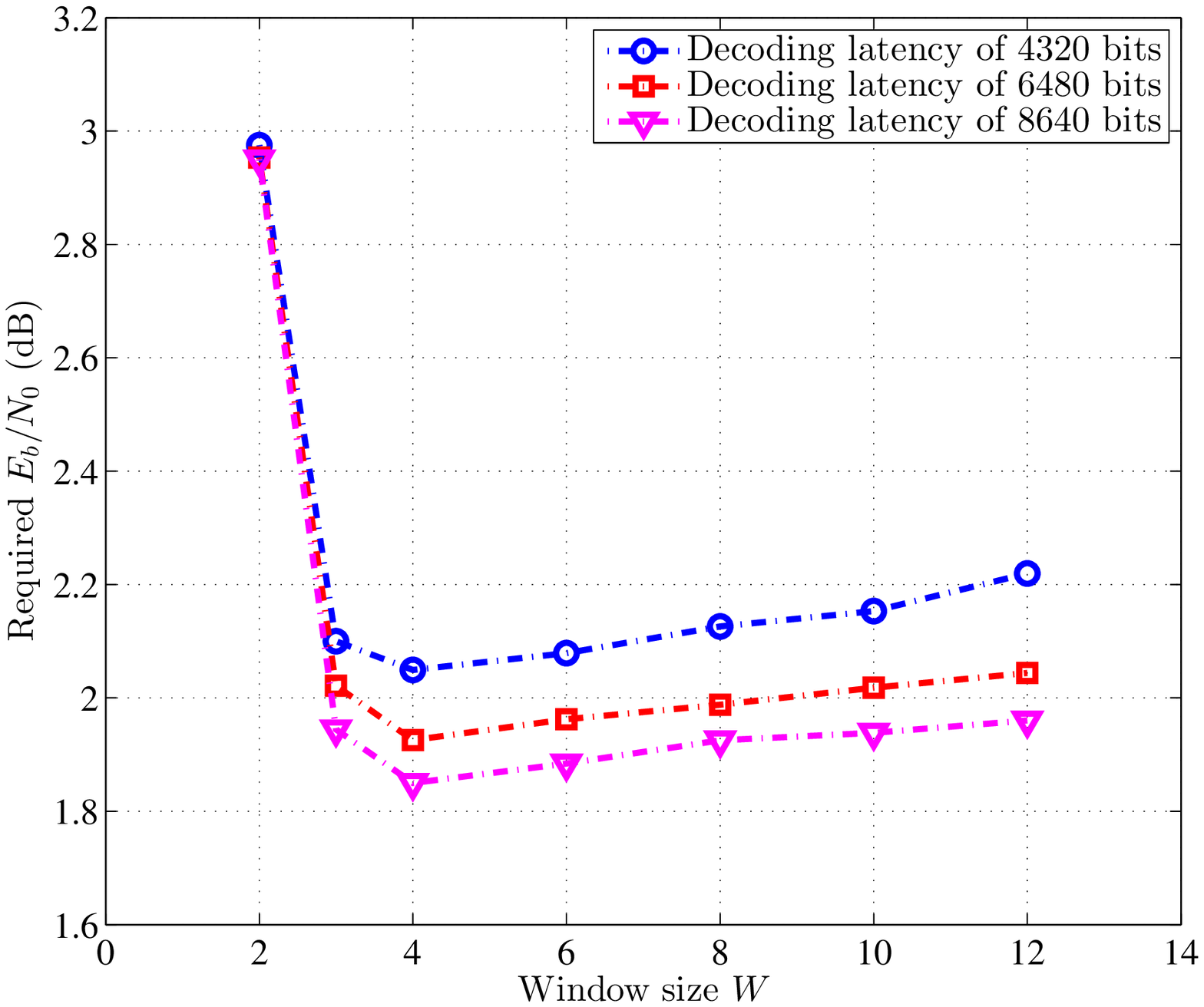}}
\subfigure[$(3,12)$-regular 8-ary SC-LDPC codes]
{
\label{Fig_3_12_ELGF8}
\includegraphics[clip, width=\figwidth]{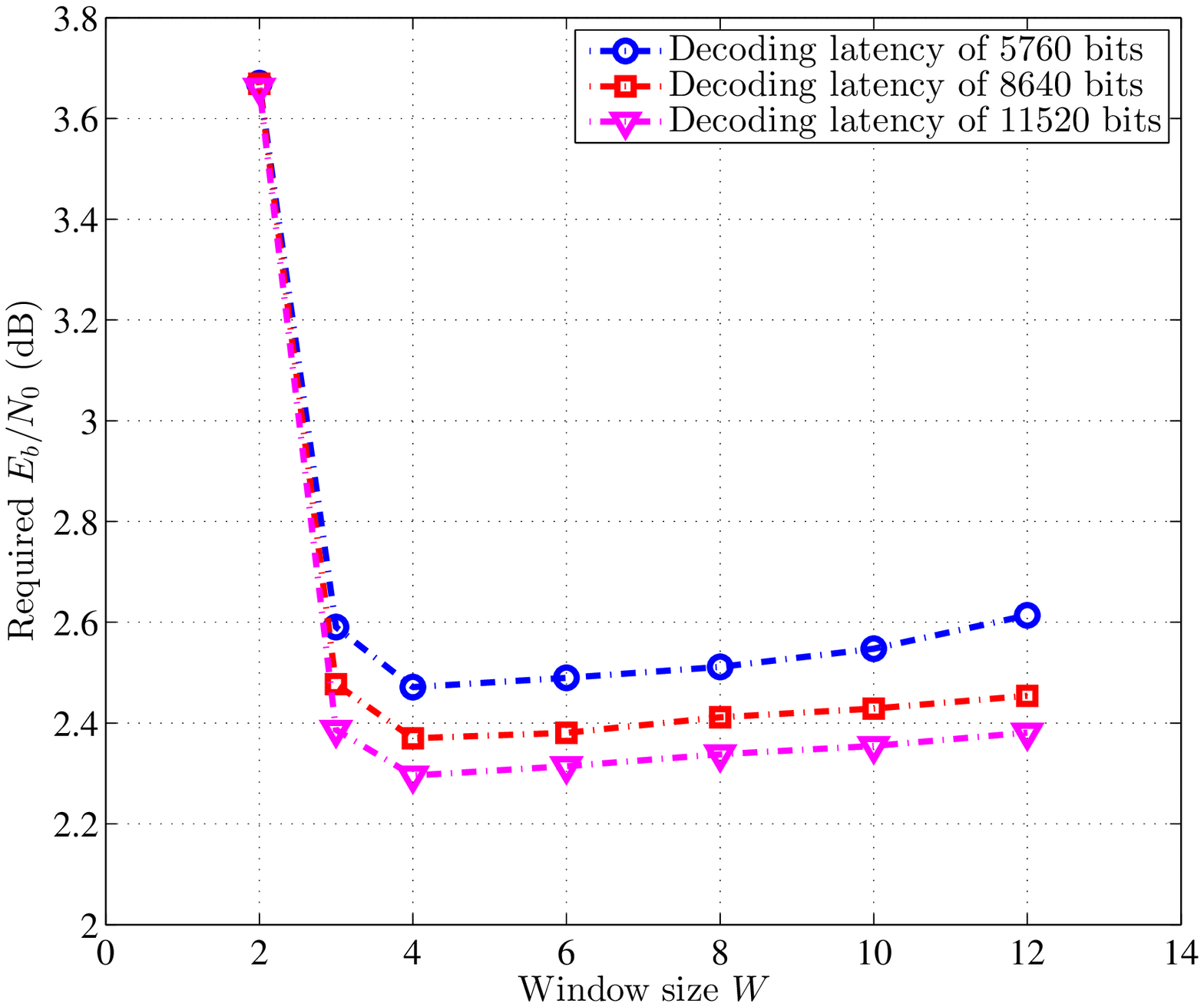}}
\caption{Required $E_b/N_0$ to achieve a BER of $10^{-5}$ for high-rate 8-ary SC-LDPC codes with different window sizes $W$ and different decoding latencies.}
\label{Fig_HighRateSameLatencyGF8}
\ifCLASSOPTIONonecolumn
\end{figure*}
\fi
\ifCLASSOPTIONtwocolumn
\end{figure}
\fi

Fig.~\ref{Fig_HighRateSameLatencyGF8} shows the $E_b/N_0$ required for the $(3,9)$-regular and $(3,12)$-regular 8-ary SC-LDPC codes to achieve a BER of $10^{-5}$ with different window sizes $W$ and different decoding latencies. We observe that the required $E_b/N_0$ for both $(3,9)$-regular and $(3,12)$-regular 8-ary SC-LDPC codes decreases dramatically until $W=4$, and then it increases gradually as $W$ increases. We therefore conclude that, for $(3,9)$-regular and $(3,12)$-regular 8-ary SC-LDPC codes, $W=4$ is a good choice for optimum performance.

\begin{table*}
\caption{Minimum $E_b/N_0$ required to achieve a BER of $10^{-5}$ for $(3,9)$-regular and $(3,12)$-regular $q$-ary LDPC-BCs and SC-LDPC codes with different field sizes}\label{table2}
  \centering
  \begin{tabular}{|c||c|c|c|c|c|c|c|c|}
  \hline
  \multirow{2}{*}{Required $\rm{E_b/N_0}$ (dB)} &\multicolumn{4}{c|}{LDPC-BC} &\multicolumn{4}{c|}{SC-LDPC~($W=4$)}\\ \cline{2-9}
  &GF($2$) &GF($4$) &GF($8$) &GF($16$) &GF($2$) &GF($4$) &GF($8$) &GF($16$) \\ \hline
  $(3,9)$ codes with latency of 4320 bits &2.4 &2.3 &2.3 &2.4 &2.5 &2.2 &2.0 &2.0 \\\hline
  $(3,9)$ codes with latency of 8640 bits &2.2 &2.1 &2.1 &2.2 &2.2 &1.9 &1.8 &1.8 \\\hline \hline
  $(3,12)$ codes with latency of 4608 bits &2.8 &2.7 &2.7 &2.8 &3.0 &2.7 &2.6 &2.5 \\\hline
  $(3,12)$ codes with latency of 9216 bits &2.7 &2.6 &2.6 &2.7 &2.7 &2.4 &2.3 &2.3 \\\hline
\end{tabular}
\end{table*}

Table~\ref{table2} shows the minimum $E_b/N_0$ required to achieve a BER of $10^{-5}$ for some $(3,9)$-regular and $(3,12)$-regular $q$-ary LDPC-BCs and SC-LDPC codes with different field sizes. Similar to the $(3,6)$-regular case, it is observed that both $(3,9)$-regular and $(3,12)$-regular non-binary SC-LDPC codes outperform both binary and non-binary LDPC-BCs and binary SC-LDPC codes for fixed decoding latency, and in general, in contrast to $q$-ary LDPC-BCs, the required $E_b/N_0$ for $q$-ary SC-LDPC codes to achieve a BER of $10^{-5}$ decreases as we increase the field size $q$. This is again consistent with results obtained for the iterative decoding thresholds in~\cite{Wei13IT}, where it is shown that, for increasing $q$, the thresholds of both $(3,9)$-regular and $(3,12)$-regular $q$-ary SC-LDPC codes approach capacity, but those of both $(3,9)$-regular and $(3,12)$-regular $q$-ary LDPC-BCs diverge from capacity. Finally, note that the minimum $E_b/N_0$ required to achieve a BER of $10^{-5}$ for both $(3,9)$-regular and $(3,12)$-regular binary SC-LDPC codes is not less than for binary LDPC-BCs for the~(relatively low) latencies considered, which is again due to the error floor effect of binary SC-LDPC codes with short constraint lengths.

\section{A Computational Complexity Comparison}\label{sec:Complexity}
In~\cite{pusane2011deriving}, the authors investigated the cost of the convolutional gain of binary SC-LDPC codes compared to binary LDPC-BCs in terms of several aspects~(computational complexity, processor complexity, decoder memory requirements, and decoding latency) of the pipeline decoder architecture. In this section, we will compare the computational complexity of $q$-ary SC-LDPC codes to $q$-ary LDPC-BCs under certain assumptions, i.e., equal decoding latency or equal decoding performance.

As stated in~\cite{Barnault03}, for $q$-ary LDPC codes implemented with the FFT-QSPA, the computational complexity per iteration at a check node is $\mathcal{O}(qm)$, while that at a variable node is $\mathcal{O}(q)$. Let $I_{\rm BC}$ denote the average number of iterations performed to decode the entire block for LDPC-BCs, and let $I_{\rm SC}$ denote the average number of iterations performed to decode the target symbols in a window for SC-LDPC codes at a particular time instant. For a $(d_v,d_c)$-regular LDPC-BC with design rate $R=\frac{d_c-d_v}{d_c}$, the computational complexity per block is then given by
\ifCLASSOPTIONonecolumn
\begin{equation}\label{BC_Complexity}
    \mathcal{O}\left(\frac{T_{\rm BC}}{m} d_v q + \frac{T_{\rm BC}}{m} \left(1-R\right) d_c qm\right)I_{\rm BC}=
    \mathcal{O}\left( \left(\frac{d_v}{m} + d_v\right)q T_{\rm BC} \right)I_{\rm BC}.
\end{equation}
\fi
\ifCLASSOPTIONtwocolumn
\begin{flalign}
&\mathcal{O}\left(\frac{T_{\rm BC}}{m} d_v q + \frac{T_{\rm BC}}{m} \left(1-R\right) d_c qm\right)I_{\rm BC}= \nonumber\\
&\mathcal{O}\left( \left(\frac{d_v}{m} + d_v\right)q T_{\rm BC} \right)I_{\rm BC}.
\end{flalign}
\fi
Thus, the computational complexity per decoded bit for a $(d_v,d_c)$-regular LDPC-BC is
\begin{equation}\label{BC_Complexity_bit}
    \mathcal{O}\left( \left(\frac{d_v}{m} + d_v\right)q \right) I_{\rm BC}.
\end{equation}

For a $(d_v,d_c)$-regular SC-LDPC code, for simplicity we consider the section of the graph covered by the window to be $(d_v,d_c)$-regular, even though the check nodes at the beginning of the window and the variable nodes at the end of the window have lower degrees. Thus the computational complexity per window is~(approximately) given by
\begin{equation}\label{SC_Complexity}
    \mathcal{O}\left( \left(\frac{d_v}{m} + d_v\right)q T_{\rm SC} \right) I_{\rm SC}.
\end{equation}
Note that the number of decoded~(target) bits for the window decoder at each time instant is $T_{\rm SC}/W$, and thus the computational complexity per decoded bit for a $(d_v,d_c)$-regular SC-LDPC code is
\begin{equation}\label{SC_Complexity_bit}
    \frac{\mathcal{O}\left( \left(\frac{d_v}{m} + d_v\right)q T_{\rm SC} \right) I_{\rm SC}}{T_{\rm SC}/W}=\mathcal{O}\left( \left(\frac{d_v}{m} + d_v\right)q \right) W I_{\rm SC}.
\end{equation}
By comparing~(\ref{BC_Complexity_bit}) and~(\ref{SC_Complexity_bit}), we see that if $I_{\rm BC}=W I_{\rm SC}$, $(d_v,d_c)$-regular LDPC-BCs and $(d_v,d_c)$-regular SC-LDPC codes with the same field size $q$ will have the same computational complexity.

In the remainder of this section we restrict our attention to $(3,6)$-regular LDPC codes; however, similar behavior has also been observed for other $(d_v,d_c)$-regular LDPC codes. For the SC-LDPC codes, the window size is set to $W=6$.

\subsection{Equal Decoding Latency}

\begin{table*}
\caption{Average number of iterations $I_{\rm BC}$ and $I_{\rm SC}$ of $(3,6)$-regular $q$-ary LDPC-BCs and $(3,6)$-regular $q$-ary SC-LDPC codes with different field sizes and decoding latencies of 4608, 6912, and 13824 bits}\label{table3}
  \centering
  \begin{tabular}{|c||c|c|c|c|c|c|c|c|}
  \hline
  \multirow{2}{*}{Average number of iterations} &\multicolumn{4}{c|}{$I_{\rm BC}$} &\multicolumn{4}{c|}{$I_{\rm SC}$~($W=6$)}\\ \cline{2-9}
  &GF($2$) &GF($4$) &GF($8$) &GF($16$) &GF($2$) &GF($4$) &GF($8$) &GF($16$) \\ \hline
  Latency of 4608 bits &13.8 &12.3 &11.1 &10.1 &3.3 &3.2 &3.0 &2.8\\\hline
  Latency of 6912 bits &15.6 &14.1 &12.6 &11.4 &3.9 &3.7 &3.4 &3.1\\\hline
  Latency of 13824 bits &19.0 &16.9 &15.5 &13.1 &5.3 &4.8 &4.4 &4.1\\\hline
\end{tabular}
\end{table*}

\begin{figure}
    \center
    \includegraphics[clip, width=\figwidth]{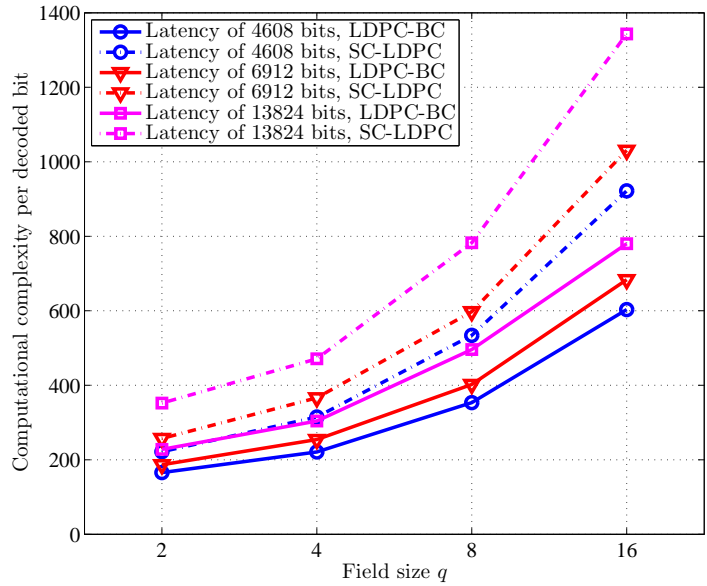}
    \caption{Computational complexity per decoded bit of $(3,6)$-regular $q$-ary SC-LDPC codes and $(3,6)$-regular $q$-ary LDPC-BCs as a function of field size $q$ with decoding latencies of 4608, 6912, and 13824 bits. The window size of the sliding window decoder for the SC-LDPC codes is $W=6$. Solid curves represent LDPC-BCs, while dotted curves represent SC-LDPC codes.}
    \label{Fig_EqualLatencyComplexity}
\end{figure}
In this subsection, we compare the computational complexity of $q$-ary SC-LDPC codes and $q$-ary LDPC-BCs under an equal decoding latency assumption. Table~\ref{table3} shows the average number of iterations $I_{\rm BC}$ and $I_{\rm SC}$ of $(3,6)$-regular $q$-ary LDPC-BCs and $(3,6)$-regular $q$-ary SC-LDPC codes with decoding latencies of 4608, 6912, and 13824 bits. We observe that $I_{\rm BC}$ for LDPC-BCs is significantly higher than $I_{\rm SC}$ for SC-LDPC codes with the same field size $q$. This results from the fact that, for a given latency, one must decode $W$ times as many target symbols for an LDPC-BC as for an SC-LDPC code. We also note that the required number of iterations for both LDPC-BCs and SC-LDPC codes decreases with $q$; however, the overall complexity increases~(see~Fig.~\ref{Fig_EqualLatencyComplexity}) because the complexity per iteration is higher.

The resulting computational complexity per decoded bit of $(3,6)$-regular $q$-ary SC-LDPC codes and $(3,6)$-regular $q$-ary LDPC-BCs with decoding latencies of 4608, 6912, and 13824 bits is shown in Fig.~\ref{Fig_EqualLatencyComplexity}.\footnote{The computational complexity results for SC-LDPC codes shown in Figs.~\ref{Fig_EqualLatencyComplexity} and~\ref{Fig_EqualPerformanceComplexity} are calculated exactly for each case, such that the slight node irregularity at the beginning and end of the window is incorporated. The resulting complexity is thus slightly lower than would be estimated using~(\ref{SC_Complexity_bit}), where the graph is assumed to be regular within a window.} We observe that the computational complexity of both SC-LDPC codes and LDPC-BCs increases exponentially with field size $q$, and the complexity of SC-LDPC codes is generally about 35\% higher than that of LDPC-BCs with the same field size $q$. From Fig.~\ref{Fig_EqualLatencyComplexity}, we also observe that the complexity of binary SC-LDPC codes is about 10\% higher than that of 4-ary LDPC-BCs, and that the complexity of 4-ary SC-LDPC codes is about 80\% higher than that of binary LDPC-BCs. However, under the equal latency assumption, binary SC-LDPC codes gain about 0.3 dB compared to 4-ary LDPC-BCs, and 4-ary SC-LDPC codes gain about 0.4 dB compared to binary LDPC-BCs~(see Table~\ref{table1} in Section~\ref{subsec:3_6}). So, even though complexity is higher for the SC-LDPC codes, the performance improvement is significant and, moreover, it is not possible to achieve this improved performance by increasing the complexity of the LDPC-BCs, i.e., allowing further iterations for LDPC-BCs will not decrease the gap in performance. We therefore conclude that, for a given latency, SC-LDPC codes provide attractive and flexible trade-offs between BER performance and computational complexity that are not available with LDPC-BCs.

\subsection{Equal Decoding Performance}

\begin{figure}
    \center
    \includegraphics[clip, width=\figwidth]{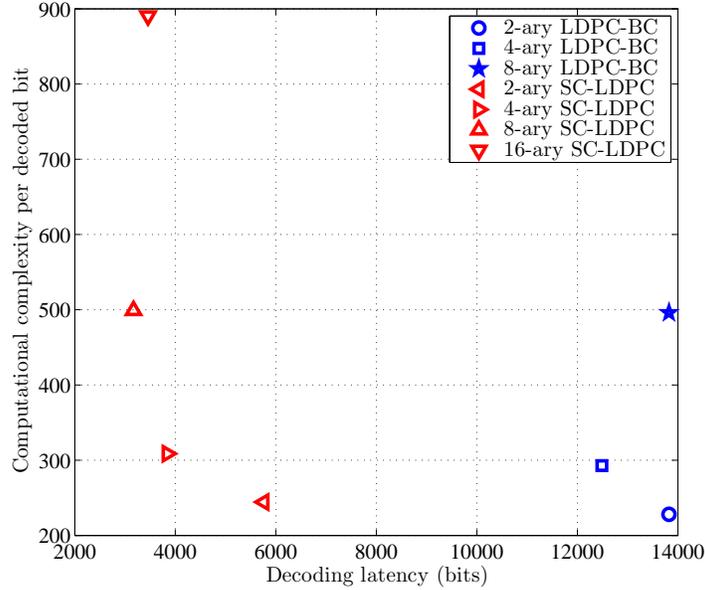}
    \caption{Computational complexity per decoded bit of $(3,6)$-regular $q$-ary SC-LDPC codes and $(3,6)$-regular $q$-ary LDPC-BCs requiring $E_b/N_0=1.5$ dB to achieve a BER of $10^{-5}$. The window size of the sliding window decoder is $W=6$.}
    \label{Fig_EqualPerformanceComplexity}
\end{figure}

In this subsection, we compare the computational complexity of $q$-ary SC-LDPC codes and $q$-ary LDPC-BCs under an equal decoding performance assumption. The computational complexity per decoded bit of $(3,6)$-regular $q$-ary SC-LDPC codes and $(3,6)$-regular $q$-ary LDPC-BCs requiring $E_b/N_0=1.5$ dB to achieve a BER of $10^{-5}$ is shown in Fig.~\ref{Fig_EqualPerformanceComplexity}.\footnote{The $(3,6)$-regular 16-ary LDPC-BC does not appear in the figure due to its large decoding latency and high computational complexity.} In general, we note that under an equal performance assumption, the SC-LDPC codes have approximately equal computational complexity as the LDPC-BCs for the same field size $q$, but a significantly reduced latency. For the SC-LDPC codes, the decoding latency decreases as the field size $q$ increases until $q=8$, and then it begins to increase as $q$ increases further, while the computational complexity increases gradually with increasing $q$ until $q=8$, and then it increases dramatically as $q$ increases further. This implies that, under these conditions, it is not worth using an SC-LDPC code with field size $q>8$. We observe the same trend for the LDPC-BCs, but with much larger latencies, and we note that the latency begins to increase for smaller values of $q$ than for the SC-LDPC codes. To be more specific, the decoding latency for the LDPC-BCs~(which is higher than for the SC-LDPC codes) decreases as the field size $q$ increases from $q=2$ to $q=4$, and then it increases as $q$ increases further, while the decoding complexity increases in line with the SC-LDPC codes. This implies that, under these conditions, it is not worth using an LDPC-BC with field size $q>4$.

From Fig.~\ref{Fig_EqualPerformanceComplexity}, we also observe that the computational complexity of the binary SC-LDPC code is about 15\% less than that of the 4-ary LDPC-BC, with about 55\% less latency. Finally, we observe that the computational complexity of the 4-ary SC-LDPC code is about 25\% higher than that of the binary SC-LDPC code, but with about 35\% less latency, and the complexity of the 4-ary SC-LDPC code is about 35\% higher than that of the binary LDPC-BC, but with about 70\% less latency. We therefore conclude that, for the same performance, 4-ary SC-LDPC codes provide attractive and flexible trade-offs between latency and computational complexity compared to using binary LDPC codes.

\subsection{Discussion}
\begin{itemize}
  \item If we fix decoding latency, we gain in decoding performance by using $q$-ary SC-LDPC codes, but at the cost of slightly higher computational complexity. For example, when the decoding latency is fixed, non-binary SC-LDPC codes with small field size $q$ outperform both binary and non-binary LDPC-BCs and binary SC-LDPC codes, while their computational complexity is slightly higher.
  \item If we fix decoding performance, we can reduce decoding latency by using $q$-ary SC-LDPC codes, but this comes at the cost of slightly higher computational complexity. For example, when the decoding performance is fixed, non-binary SC-LDPC codes with small field size $q$ have lower decoding latency than both binary and non-binary LDPC-BCs and binary SC-LDPC codes, while their computational complexity is slightly higher.
  \item Overall, these results imply that $(3,6)$-regular 4-ary SC-LDPC codes possess a particularly attractive combination of small decoding latency, low computational complexity, and good decoding performance.
\end{itemize}

\section{Conclusions}\label{sec:Conclusion}
In this paper, we considered a finite-length performance comparison of  protograph-based $q$-ary SC-LDPC codes and $q$-ary LDPC-BCs. We proposed a sliding window decoding algorithm with a stopping rule based on a soft BER estimate for $q$-ary SC-LDPC codes. Simulation results confirm that $(2,4)$-, $(3,6)$-, $(3,9)$-, and $(3,12)$-regular $q$-ary SC-LDPC codes achieve substantial convolutional gains compared to the underlying LDPC-BCs, where the constraint length of the SC-LDPC codes is equal to the block length of the LDPC-BCs.

We also examined the relationship between the protograph lifting factor, the decoding window size, and the BER performance of $q$-ary SC-LDPC codes for fixed decoding latency in comparison to $q$-ary LDPC-BCs. It was observed that, under an equal latency constraint, $(3,6)$-regular non-binary SC-LDPC codes outperform both binary and non-binary LDPC-BCs and binary SC-LDPC codes. Moreover, for fixed field size and latency, the decoding performance of $(3,6)$-regular $q$-ary SC-LDPC codes improves as the window size $W$ increases up to a certain point~(around $W=6$), and then it degrades slightly as $W$ increases further. Similar behavior was also observed for $(3,9)$-regular and $(3,12)$-regular $q$-ary SC-LDPC codes in comparison to their $q$-ary LDPC-BC counterparts.

Finally, we compared the computational complexity of $q$-ary SC-LDPC codes to $q$-ary LDPC-BCs under equal decoding latency and equal decoding performance assumptions. It was observed that $(3,6)$-regular 4-ary  SC-LDPC codes have a particularly attractive combination of small decoding latency, low computational complexity, and good decoding performance. An interesting future research topic to complement the work reported here would be to design the permutations and edge labels used in the construction process, rather than to select them randomly, to further improve the performance of $q$-ary SC-LDPC codes for a given decoding latency.


\ifCLASSOPTIONcaptionsoff
  \newpage
\fi


\begin{thebibliography}{10}
\bibitem{Gallager63}
R.~G. Gallager, \emph{Low-Density Parity-Check Codes}.\hskip 1em plus 0.5em
  minus 0.4em\relax Cambridge, MA: MIT Press, 1963.

\bibitem{Richardson01}
T. J. Richardson, M. A. Shokrollahi, and R. L. Urbanke, ``Design of capacity-approaching irregular low-density parity-check codes,'' \emph{IEEE Trans. Inf. Theory}, vol.~47, no.~2, pp.
  619--637, Feb. 2001.

\bibitem{Davey98}
M.~C. Davey and D.~J.~C. MacKay, ``Low-density parity-check codes over
  {GF}($q$),'' \emph{IEEE Commun. Letters}, vol.~2, pp. 165--167, June 1998.

\bibitem{Barnault03}
L. Barnault and D. Declercq, ``Fast decoding algorithm for LDPC over GF($2^q$),'' in \emph{Proc. IEEE Inf. Theory Workshop}, Paris, France, pp. 70--73, Mar. 2003.


\bibitem{Declercq07}
D.~Declercq and M.~Fossorier, ``Decoding algorithms for nonbinary {LDPC} codes over {GF}($q$),'' \emph{IEEE Trans.~Commun.}, vol.~55, no.~4, pp. 633--643, Apr. 2007.

\bibitem{Voicila10}
A.~Voicila, D.~Declercq, F.~Verdier, M.~Fossorier and P.~Urard, ``Low-complexity decoding for non-binary LDPC codes in high order fields,'' \emph{IEEE Trans.~Commun.}, vol.~58, no.~5, pp. 1365--1375, May 2010.

\bibitem{Ma12}
X.~Ma, K.~Zhang, H.~Chen, and B.~Bai, ``Low complexity {X-EMS} algorithms for nonbinary {LDPC} codes,'' \emph{IEEE Trans.~Commun.}, vol.~60, no.~1, pp. 9--13, Jan. 2012.

\bibitem{poulliat08}
C.~Poulliat, M.~Fossorier, and D.~Declercq, ``Design of regular (2,$d_c$)-LDPC codes over {GF}($q$) using their binary images,'' \emph{IEEE Trans.~Commun.}, vol.~56, no.~10, pp. 1626--1635, Oct. 2008.

\bibitem{Zeng08}
L.~Zeng, L.~Lan, Y.~Y.~Tai, S.~Song, S.~Lin, and K. Abdel-Ghaffar, ``Constructions of nonbinary quasi-cyclic LDPC codes: A finite field approach,'' \emph{IEEE Trans.~Commun.}, vol.~56, no.~4, pp. 545--554, Apr. 2008.

\bibitem{Zhao13}
S.~Zhao, X.~Ma, X.~Zhang, and B.~Bai, ``A class of nonbinary LDPC codes with fast encoding and decoding algorithms,'' \emph{IEEE Trans.~Commun.}, vol.~61, no.~1, pp. 1--6, Jan. 2013.

\bibitem{Dolecek13}
L.~Dolecek, D.~Divsalar, Y.~Sun, and B.~Amiri, ``Non-binary protograph-based {LDPC} codes: Enumerators, analysis, and designs.'' \emph{IEEE Trans. Inf. Theory}, vol.~60, no.~7, pp. 3913--3941, July 2014.

\bibitem{jimenez99}
A.~J. Felstr\"{o}m and K.~Sh. Zigangirov, ``Time-varying periodic convolutional codes with low-density parity-check matrix,'' \emph{IEEE Trans. Inf. Theory}, vol.~45, no.~6, pp. 2181--2191, Sept. 1999.

\bibitem{lentmaier10}
M.~Lentmaier, A.~Sridharan, D.~J. Costello,~Jr., and K.~Sh. Zigangirov, ``Iterative decoding threshold analysis for {LDPC} convolutional codes,'' \emph{IEEE Trans. Inf. Theory}, vol.~56, no.~10, pp. 5274--5289, Oct. 2010.

\bibitem{kudekar11}
S.~Kudekar, T.~J.~Richardson, and R. L. Urbanke, ``Threshold saturation via spatial coupling: Why convolutional LDPC ensembles perform so well over the BEC,'' \emph{IEEE Trans. Inf. Theory}, vol.~57, no.~4, pp. 803--834, Feb. 2011.

\bibitem{kudekar13}
S.~Kudekar, T.~J.~Richardson, and R. L. Urbanke, ``Spatially coupled ensembles universally achieve capacity under belief propagation,'' \emph{IEEE Trans. Inf. Theory}, vol.~59, no.~12, pp. 7761--7813, Dec. 2013.

\bibitem{Ali08}
A.~E. Pusane, A.~J. Felstr\"{o}m, A.~Sridharan, M.~Lentmaier, K.~Sh. Zigangirov, and D.~J. Costello,~Jr., ``{Implementation aspects of LDPC convolutional codes},'' \emph{IEEE Trans.~Commun.}, vol.~56, no.~7, pp. 1060--1069, July 2008.

\bibitem{iyengar2012windowed}
A.~R. Iyengar, M.~Papaleo, P.~H. Siegel, J.~K. Wolf, A.~Vanelli-Coralli, and G.~E. Corazza, ``Windowed decoding of protograph-based LDPC convolutional codes over erasure channels,'' \emph{IEEE Trans. Inf. Theory}, vol.~58, no.~4, pp. 2303--2320, Apr. 2012.

\bibitem{Uchikawa11}
H.~Uchikawa, K.~Kasai, and K.~Sakaniwa, ``{Design and performance of rate-compatible non-binary LDPC convolutional codes}.'' \emph{IEICE Transactions on Fundamentals of Electronics, Communications and Computer Sciences}, vol.~94, no.~11, pp.~2135--2143, Nov. 2011. [online]. Available: {http://arxiv.org/abs/1010.0060}

\bibitem{Piemontese13}
I. Andriyanova and A. Graell i Amat, ``{Threshold saturation for nonbinary SC-LDPC codes on the binary erasure channel},'' submitted to \emph{IEEE Trans. Inf. Theory}, 2013. [online]. Available: {http://arxiv.org/abs/1311.2003}

\bibitem{Thorpe03}
J.~Thorpe, ``{Low-density parity-check (LDPC) codes constructed from protographs},'' JPL INP Progress Report 42-154, Aug. 2003.

\bibitem{Wei13}
L.~Wei, T. Koike-Akino, D. G. M. Mitchell, T. E. Fuja, and D.~J.~Costello,~Jr., ``Threshold analysis of non-binary spatially-coupled codes with windowed decoding,'' in \emph{Proc.~IEEE Int. Symp. on Inf. Theory}, Honolulu, HI, July 2014.

\bibitem{Wei13IT}
L.~Wei, D. G. M. Mitchell, T. E. Fuja, and D.~J.~Costello,~Jr., ``Design of spatially-coupled {LDPC} codes over GF($q$) with windowed decoding,'' submitted to \emph{IEEE Trans. Inf. Theory}, 2014.

\bibitem{lentmaier09}
M.~Lentmaier, G.~P. Fettweis, K.~Sh. Zigangirov, and D.~J.~Costello,~Jr., ``Approaching capacity with asymptotically regular {LDPC} codes,'' in \emph{Proc. Inf. Theory and Appl. Workshop,} San Diego, CA, pp. 173--177, Feb. 2009.

\bibitem{lentmaier11}
M.~Lentmaier, M.~M.~Prenda, and G.~P. Fettweis, ``{Efficient message passing scheduling for terminated LDPC convolutional codes},'' in \emph{Proc.~IEEE Int. Symp. on Inf. Theory}, St. Petersburg, Russia, pp. 1826--1830, Aug. 2011.

\bibitem{lentmaier12}
N.~ul~Hassan, M.~Lentmaier, and G.~P. Fettweis, ``{Comparison of LDPC block and LDPC convolutional codes based on their decoding latency},'' in \emph{Proc. Int. Symp. Turbo Codes Iterative Inf. Process.}, Gothenburg, Sweden, pp. 225--229, Aug. 2012.

\bibitem{bates2008low}
S.~Bates, Z.~Chen, L.~Gunthorpe, A.~E. Pusane, K.~Sh. Zigangirov, and D.~J. Costello, Jr., ``{A low-cost serial decoder architecture for low-density parity-check convolutional codes},'' \emph{IEEE Trans. Circuits Syst. I, Reg. Papers}, vol.~55, no.~7, pp. 1967--1976, Aug. 2008.

\bibitem{pusane2011deriving}
A.~E. Pusane, R. Smarandache, P.~O. Vontobel, and D.~J. Costello, Jr., ``Deriving good LDPC convolutional codes from LDPC block codes,'' \emph{IEEE Trans. Inf. Theory}, vol.~57, no.~2, pp. 835--857, Feb. 2011.

\bibitem{Hassan12}
N.~ul~Hassan, A.~E. Pusane, M.~Lentmaier, G.~P. Fettweis, and D.~J. Costello, Jr., ``{Reduced complexity window decoding schedules for coupled LDPC codes},'' in \emph{Proc. IEEE Inf. Theory Workshop}, Lausanne, Switzerland, pp. 20--24, Sept. 2012.

\bibitem{Hassan13}
N.~ul~Hassan, A.~E. Pusane, M.~Lentmaier, G.~P. Fettweis, and D.~J. Costello, Jr., ``{Non-uniform windowed decoding schedules for spatially coupled codes},'' in \emph{Proc. IEEE Global Commun. Conf.}, Atlanta, GA, Dec. 2013.


\end{thebibliography}
\end{document}